\documentclass[twocolumn]{aastex631}

\usepackage{multirow}

\usepackage{mathtools}

\renewcommand{\sout}{\bgroup \color{red} \ULdepth=-.5ex \ULset}

\graphicspath{{./}{figures/}}
\begin{document}

\title{High density symmetry energy: A key to the solution of the hyperon puzzle}

\author[0000-0001-6899-6009]{Jun-Ting Ye}
\affiliation{School of Physics and Astronomy, Shanghai Key Laboratory for
Particle Physics and Cosmology, and Key Laboratory for Particle Astrophysics and Cosmology (MOE),
Shanghai Jiao Tong University, Shanghai 200240, China}

\author[0000-0002-6465-6186]{Rui Wang}
\affiliation{Istituto Nazionale di Fisica Nucleare (INFN), Sezione di Catania, I-95123 Catania, Italy}

\author[0009-0008-7129-8197]{Si-Pei Wang}
\affiliation{School of Physics and Astronomy, Shanghai Key Laboratory for
Particle Physics and Cosmology, and Key Laboratory for Particle Astrophysics and Cosmology (MOE),
Shanghai Jiao Tong University, Shanghai 200240, China}

\author[0000-0002-7444-0629]{Lie-Wen Chen}
\affiliation{School of Physics and Astronomy, Shanghai Key Laboratory for
Particle Physics and Cosmology, and Key Laboratory for Particle Astrophysics and Cosmology (MOE),
Shanghai Jiao Tong University, Shanghai 200240, China}

\correspondingauthor{Lie-Wen Chen}
\email{lwchen@sjtu.edu.cn}

\begin{abstract}
The recently developed nuclear effective interaction based on the so-called N3LO
Skyrme pseudopotential is extended to include the
hyperon-nucleon and hyperon-hyperon interactions by assuming the similar density, momentum, and isospin
dependence as for the nucleon-nucleon interaction.
The parameters in these interactions are determined from either experimental information if any or chiral effective field theory or lattice QCD calculations
of the hyperon potentials in nuclear matter around nuclear saturation density $\rho_0$.
We find that
varying the high density behavior of the symmetry energy $E_{\rm sym}(\rho)$ can significantly change the critical density for hyperon appearance in the neutron stars and thus the maximum mass $M_{\rm TOV}$ of static hyperon stars.
In particular,
a symmetry energy
which is soft around $2-3\rho_0$ but stiff above about $4\rho_0$, can lead to $M_{\rm TOV} \gtrsim 2M_\odot$ for hyperon stars and simultaneously be compatible with
(1) the constraints on the equation of state of symmetric nuclear matter at suprasaturation densities obtained from flow data in heavy-ion collisions;
(2) the microscopic calculations of the equation of state for pure neutron matter;
(3) the star tidal deformability extracted from gravitational wave signal GW170817;
(4) the mass-radius relations of PSR J0030+0451, PSR J0740+6620 and PSR J0437-4715 measured from NICER;
(5) the observation of the unusually low mass and small radius in the central compact object of HESS J1731-347.
Furthermore, the squared sound speed of the hyperon star matter naturally displays a strong peak structure around baryon density of $3-4\rho_0$, consistent with the model-independent analysis on the multimessenger data.
Our results suggest that
the high density symmetry energy could be a key to the solution of the hyperon puzzle in neutron star physics.
\end{abstract}

%----------------------------------------------------------------------------------------------------
\section{Introduction}
In the dense cores of neutron stars (NSs), the chemical potentials of nucleons may reach sufficiently high levels to enable their $\beta$ decay into hyperons.
Neutron stars with such interiors, consisting not only of nucleons and leptons but also of hyperons, are referred to as hyperon stars (HSs).
The presence of hyperons releases Fermi pressure and renders the equation of state (EOS) of NS matter softer, consequently resulting in that the maximum mass of NSs might not align with astrophysical observations of massive NSs with mass larger than about $2M_{\odot}$~\citep{Antoniadis:2013pzd,NANOGrav:2019jur,Fonseca:2021wxt}.
This phenomenon is commonly known as ``hyperon puzzle"~\citep{Chatterjee:2015pua,Vidana:2018bdi,Tolos:2020aln,Bombaci:2021ffs,Burgio:2021vgk,Vidana:2022tlx,Tolos:2024xyi}.

The study of the hyperon puzzle is closely linked to the EOS of dense nuclear matter, which reflects the fundamental properties of effective nuclear interactions.
Experimentally,
analyses on the data from the giant monopole resonance in finite nuclei~\citep{Youngblood:1999zza,Li:2007bp,Li:2022suc,Shlomo:2006ole,Garg:2018uam} as well as the measurements of collective flow~\citep{Danielewicz:2002pu,LeFevre:2015paj} and kaon production~\citep{Aichelin:1985rbt,Fuchs:2000kp,Hartnack:2005tr,Fuchs:2005zg} in high energy heavy-ion collisions~(HICs) have already provided valuable constraints on the EOS of symmetric nuclear matter~(SNM) from nuclear saturation density $\rho_0$ to suprasaturation density of about $5\rho_0$.
On the other hand,
for the isospin-dependent part of nuclear matter EOS, characterized by the so-called nuclear symmetry energy $E_{\rm sym}(\rho)$,
its suprasaturation density behavior is still among the most uncertain properties of nuclear matter~\citep{Li:2008gp,Sorensen:2023zkk}, although the symmetry energy at densities below and around $\rho_0$ has been reasonably well-constrained through nuclear structure probes~\citep{Chen:2005ti,Centelles:2008vu,Chen:2010qx,Agrawal:2012pq,Zhang:2013wna,Brown:2013mga,Roca-Maza:2012uor,Zhang:2014yfa,Danielewicz:2013upa,Danielewicz:2016bgb,Xu:2020xib,Qiu:2023kfu} and observables in HICs at energies less than about $100$~MeV/nucleon~\citep{Chen:2004si,Li:2005jy,Kowalski:2006ju,Shetty:2005qp,Tsang:2008fd,Wada:2011qm,Morfouace:2019jky,Zhang:2020azr}.
In recent years, significant progress has been made to constrain the high density behavior of the symmetry energy by using the multimessenger data, especially the gravitational wave observation in binary neutron star merger as well as the simultaneous determination of the mass and radius of NSs by NICER, but the uncertainty of the constraints is still very large~\citep{FiorellaBurgio:2018dga,Zhang:2018bwq,Zhou:2019omw,Zhou:2019sci,Li:2021thg,Krastev:2021reh,Yue:2021yfx,Koehn:2024set}.
Besides adhering to the constraints imposed by terrestrial experiments and astrophysical observations, it is interesting to mention that the systematics of the density dependence of the symmetry energy based on a large samples of theoretical models can also provide estimate on the suprasaturation density behavior of the symmetry energy from their precise knowledge around saturation density~\citep{Chen:2011ib,Chen:2015gba}.

While extensive research has focused on understanding the nucleon-nucleon~(NN) interaction, comparatively less attention has been given to investigating the hyperon-nucleon~(YN) and the hyperon-hyperon~(YY) interactions.
Experimentally,
the available scattering data includes several hundred measurements for $\Lambda N$~\citep{Alexander:1968acu,Sechi-Zorn:1968mao,Kadyk:1971tc} and $\Sigma N$~\citep{Engelmann:1966npz,Eisele:1971mk}, but there is a scarcity of information regarding $\Xi N$.
Furthermore, over fourty single $\Lambda$-hypernuclei as well as some double~$\Lambda$ \citep{Danysz:1963zz,Prowse:1966nz,Takahashi:2001nm} and single $\Xi$~\citep{Vidana:2018bdi} hypernuclei have been discovered.
However, experimental data for YY interactions remains missing.
Theoretically,
various models have been developed by extending the nucleonic interaction to describe hyperonic interactions.
These models can be roughly classified into phenomenological and microscopic approaches.
Phenomenological methods, such as non-relativistic Skyrme interaction~\citep{Skyrme:1959zz} and relativistic mean-field~(RMF) theory~\citep{Walecka:1985my,Serot:1997xg}, are widely employed in this context.
The microscopic calculations involve starting with realistic interactions and constructing them using meson-exchange theory~\citep{Holzenkamp:1989tq,Stoks:1999bz,Rijken:1998yy}, quark models~\citep{Fujiwara:2006yh}, chiral effective field theory~($\chi$EFT)~\citep{Weinberg:1990rz,Polinder:2006zh,Haidenbauer:2013oca}, and Lattice Quantum Chromodynamics~(LQCD)~\citep{Ishii:2006ec,Beane:2012ey,Nemura:2017vjc}.
Many-body problems are then solved through techniques like the Brueckner-Hartree-Fock~(BHF) approximation~\citep{Baldo:1999rq,Vidana:2000ew}.

Currently, several potential solutions to the hyperon puzzle have been proposed \citep{Vidana:2018bdi,Tolos:2024xyi}.
The primary objective of these approaches is to enhance the stiffness of the hyperonic matter EOS in order to ensure that the maximum mass of HSs remains sufficiently large.
Strengthening YN and YY interactions~\citep{Bednarek:2011gd,Weissenborn:2011ut,Jiang:2012hy,Gomes:2014aka,Maslov:2015msa,Li:2018jvz,Sun:2022yor,Wei:2024yda} or introducing more repulsive three-body forces~\citep{Tsubakihara:2012ic,Lonardoni:2014bwa,Logoteta:2019utx,Gerstung:2020ktv} involving hyperons could effectively stiffen the EOS for hyperonic matter.
Additionally, it is also plausible to consider a phase transition to deconfined quark matter with stiff EOS \citep{Ozel:2010bz,Weissenborn:2011qu,Zdunik:2012dj} occurring prior to the appearance of hyperons.
Furthermore, there is consideration regarding the inclusion of $\Delta$ baryons~\citep{Drago:2014oja,Li:2018qaw} or kaon condensate \citep{Glendenning:1998zx,Muto:2007iz}, which may delay the onset of hyperons until higher densities are reached.
Moreover, accounting for momentum dependent in-medium potentials for hyperons~\citep{Chorozidou:2024gyy} is of significant importance as well.
Some unconventional methods such as modifying gravity~\citep{Astashenok:2014pua} have even been proposed.
For the moment, nevertheless, resolution to the hyperon puzzle remains elusive.

In this work, we propose that the high density symmetry energy could be a key to solve the hyperon puzzle.
We first extend the recently developed N3LO Skyrme pseudopotential~\citep{Wang:2018yce,Wang:2023zcj} to describe the interactions of octet baryons by assuming that the YN and YY interactions have similar density, momentum and isospin dependence as the NN interaction with some scaling parameters.
The scaling parameters are then adjusted to fit the single-hyperon potentials in SNM and pure neutron matter~(PNM) obtained from experiments if any, $\mathrm{\chi EFT}$~\citep{Petschauer:2015nea} or LQCD~\citep{Inoue:2018axd}.
By varying the high density behavior of the symmetry energy, we find that the symmetry energy can significantly change the critical density of hyperon appearance in NSs, and a symmetry energy soft at intermediate density but stiff at high density can support a massive HS with mass larger than $2.0M_\odot$ and simultaneously be compatible with current popular constraints from theory, experiments and observations.

We note that
a number of studies~\citep{Ryu:2011vw,Cavagnoli:2011ft,Providencia:2012rx,Providencia:2013dsa,Bizarro:2015wxa,Providencia:2018ywl,Choi:2020eun,Ghosh:2022lam,Thapa:2021syu,Kumar:2023qcs} have been devoted to exploring the symmetry energy effects on the properties of HSs, essentially via varying the slope parameter $L$ with the high density symmetry energy changed automatically according to the adopted energy density functional. In the present work, we investigate the high density symmetry energy effects through varying independently the higher-order symmetry energy parameters, namely, the curvature parameter $K_{\rm sym}$ and the skewness parameter $J_{\rm sym}$, while keeping the $L$ parameter unchanged in our energy density functional.

This article is organized as follows. In Section~\ref{Sec:Model}, we mainly describe the model and method used in extending the recently developed N3LO Skyrme pseudopotential to include the YN and YY interactions.
We then present in Section~\ref{Sec:Result} the results and discussions on how the high density symmetry energy may influence the properties of NSs and HSs.
Finally, the conclusion and outlook are given in Section~\ref{Sec:Summary}.

%-------------------------------------------------------------------------------------------------------------------
\section{Model and method}
\label{Sec:Model}
\subsection{Nuclear matter EOS and its characteristic parameters}
The EOS of nuclear matter, which is defined as the binding energy per nucleon, can be expressed as
\begin{equation}
\label{EOS}
\begin{aligned}
    E(\rho,\delta)=E_0(\rho)+E_{\rm sym}(\rho)\delta^2+{\cal O}(\delta^4),
\end{aligned}
\end{equation}
where $\rho=\rho_{\rm n}+\rho_{\rm p}$ represents the total nucleon density, which is the sum of the neutron density $\rho_{\rm n}$ and proton density $\rho_{\rm p}$; $\delta=(\rho_{\rm n}-\rho_{\rm p})/\rho$ denotes the isospin asymmetry;
$E_0(\rho)=E(\rho,\delta=0)$ represents the EOS of SNM; and the symmetry energy $E_{\rm{sym}}(\rho)$ can be obtained as
\begin{equation}
E_{\rm{sym}}(\rho) = \left.\frac{1}{2!}\frac{\partial^{2}E(\rho,\delta)}{\partial\delta^{2}}\right|_{\delta=0}.
\end{equation}
Around the saturation density $\rho_0$, the $E_0(\rho)$ can be expressed approximately in terms of its incompressibility coefficient $K_0$ and skewness parameter $J_0$ as
\begin{equation}
E_0(\rho)=E_0(\rho_0)+\frac{1}{2!}K_0\chi^2+\frac{1}{3!}J_0\chi^3+{\cal O}(\chi^4),
\end{equation}
where $\chi=(\rho-\rho_0)/(3\rho_0)$ is a dimensionless parameter representing the density deviation from the saturation density $\rho_0$.
Similarly, by expanding $E_{\rm sym}(\rho)$ around a reference density $\rho_r$ in terms of its slope parameter $L(\rho_r)$, curvature parameter $K_{\rm{sym}}(\rho_r)$ and skewness parameter $J_{\rm{sym}}(\rho_r)$, we have
\begin{eqnarray}
E_{\rm{sym}}(\rho) &=& E_{\rm{sym}}(\rho_r) + L(\rho_r) \chi_r + \frac{1}{2!}K_{\rm{sym}}(\rho_r)\chi_r^2 \notag \\ &+&\frac{1}{3!}J_{\rm{sym}}(\rho_r)\chi_r^3+\mathcal{O}(\chi_r^4),
\end{eqnarray}
where $\chi_r$ is defined as $\chi_r=(\rho-\rho_r)/(3\rho_r)$.
By setting $\rho_r = \rho_0$, one can then obtain the conventional characteristic parameters $L \equiv L(\rho_0)$, $K_{\rm{sym}} \equiv K_{\rm{sym}}(\rho_0)$ and $J_{\rm{sym}} \equiv J_{\rm{sym}}(\rho_0)$.

\subsection{N3LO Skyrme pseudopotential for baryon octet}
We generalize the recently developed extended N3LO Skyrme psuedopotential~\citep{Wang:2018yce,Wang:2023zcj} to include the interactions of octet baryons.
When focusing on the spin-averaged quantities as we are interested in here, there only contains the central term $V_{\mathrm{N} 3 \mathrm{LO}}^\mathrm{C}$ and the density-dependent term $V_{\mathrm{N} 1 \mathrm{LO}}^{\mathrm{DD}}$ in the extended N3LO Skyrme psuedopotential $v_{Sk}$ for nucleonic interaction~\citep{Wang:2018yce,Wang:2023zcj}, namely,
\begin{equation}
\label{eq:Vsk}
    v_{Sk}= V_{\mathrm{N} 3 \mathrm{LO}}^\mathrm{C} + V_{\mathrm{N} 1 \mathrm{LO}}^{\mathrm{DD}}.
\end{equation}
The central term is given by~\citep{Wang:2018yce,Wang:2023zcj}
\begin{widetext}
\small
\begin{equation}
\label{eq:VN3LO}
\begin{aligned}
    V_{\mathrm{N} 3 \mathrm{LO}}^\mathrm{C}  =
    & t_0\left(1+x_0 \hat{P}_\sigma\right)+t_1^{[2]}\left(1+x_1^{[2]} \hat{P}_\sigma\right) \frac{1}{2}\left(\hat{\vec{k}}^{\prime 2}+\hat{\vec{k}}^2\right)+t_2^{[2]}\left(1+x_2^{[2]} \hat{P}_\sigma\right) \hat{\vec{k}}^{\prime} \cdot \hat{\vec{k}}+t_1^{[4]}\left(1+x_1^{[4]} \hat{P}_\sigma\right)\left[\frac{1}{4}\left(\hat{\vec{k}}^{\prime 2}+\hat{\vec{k}}^2\right)^2+\left(\hat{\vec{k}}^{\prime} \cdot \hat{\vec{k}}\right)^2\right] \\
    & +t_2^{[4]}\left(1+x_2^{[4]} \hat{P}_\sigma\right)\left(\hat{\vec{k}}^{\prime} \cdot \hat{\vec{k}} \right)\left(\hat{\vec{k}}^{\prime 2}+\hat{\vec{k}}^2\right)+t_1^{[6]}\left(1+x_1^{[6]} \hat{P}_\sigma\right)\left(\hat{\vec{k}}^{\prime 2}+\hat{\vec{k}}^2\right)\left[\frac{1}{2}\left(\hat{\vec{k}}^{\prime 2}+\hat{\vec{k}}^2\right)^2+6\left(\hat{\vec{k}}^{\prime} \cdot \hat{\vec{k}}\right)^2\right] \\
    & +t_2^{[6]}\left(1+x_2^{[6]} \hat{P}_\sigma\right)\left(\hat{\vec{k}}^{\prime} \cdot \hat{\vec{k}}\right)\left[3\left(\hat{\vec{k}}^{\prime 2}+\hat{\vec{k}}^2\right)^2+4\left(\hat{\vec{k}}^{\prime} \cdot \hat{\vec{k}}\right)^2\right] ,
\end{aligned}
\end{equation}
while the density-dependent term is expressed as~\citep{Wang:2018yce,Wang:2023zcj}
\begin{equation}
\label{eq:Vdd_new}
    V^{\mathrm{DD}}_{\mathrm{N} 1 \mathrm{LO}}= \sum_{n=1,3,5} \frac{1}{6} t_{3}^{\left[ n \right] }\left(1+x_{3}^{\left[ n \right] } \hat{P}_\sigma\right) \rho^{n/3}(\vec{R}),
\end{equation}
where $\hat{P}_\sigma=(1+\hat{\vec{{\sigma}}}_1 \cdot \hat{\vec{{\sigma}}}_2)/2$ is the spin-exchange operator; $\hat{\vec{k}}=-i\left( \hat{\vec{\nabla}}_1-\hat{\vec{\nabla}}_2 \right)/2$ is the relative momentum operator; $ \hat{\vec{k}}^{\prime} $ is the conjugate operator of $ \hat{\vec{k}}$ acting on the left; $\vec{R} = \left( \vec{r}_1 + \vec{r}_2 \right)/2$; the $t_0$, $x_0$, $t_{i}^{[n]}$, $x_{i}^{[n]}$ ($n=2,4,6$ and $i=1,2$), $t_{3}^{[n]}$ and $x_{3}^{[n]}$ ($n=1,3,5$) are Skyrme parameters and thus there are totally twenty parameters.
Note in Eqs.~(\ref{eq:VN3LO}) and (\ref{eq:Vdd_new}), the factor $\hat{\delta}\left(\vec{r_1}-\vec{r_1} \right)$ is omitted for brevity.

In the mean-field approximation, the potential energy density for nucleon system with the extended N3LO Skyrme psuedopotential is given by~\citep{Wang:2018yce,Wang:2023zcj}
\begin{equation}
\label{eq:VNN_old}
\begin{aligned}
    V(\rho_n,\rho_p)=&\frac{1}{4} t_0 \left[ \left( 2+x_0 \right) \rho^2 - \left( 2 x_0+1 \right) \sum_{\tau=n, p} \rho_\tau^2\right]+
    \sum_{n=1,3,5} \frac{1}{24} t_3^{[n]} \left[ \left(2+x_3^{[n]}\right) \rho^2 -\left(2 x_3^{[n]}+1\right) \sum_{\tau=n, p} \rho_\tau^2 \right] \rho^{n / 3}\\
    &+\sum_{k=2,4,6} \Bigg\{ \frac{C^{[k]}}{16 \hbar^k (1+\delta_{k4})} \int d^3 p d^3 p^{\prime}\left(\vec{p}-\vec{p}^{\, \prime}\right)^k f(\vec{r}, \vec{p}) f\left(\vec{r}, \vec{p}^{\, \prime}\right)\\
    &+\frac{D^{[k]}}{16 \hbar^k (1+\delta_{k4})} \sum_{\tau=n, p} \int d^3 p d^3 p^{\prime}\left(\vec{p}-\vec{p}^{\, \prime}\right)^k f_{\tau}(\vec{r}, \vec{p}) f_{\tau}\left(\vec{r}, \vec{p}^{\, \prime}\right) \\
    &+\sum_{m=0}^{k} (-1)^{m}{\dbinom{k}{m}} \bigg[ \frac{E^{[k]}}{16 (1+\delta_{k4})} {\nabla^m}{\rho}{\nabla^{k-m}}{\rho}+\frac{F^{[k]}}{16 (1+\delta_{k4})}  \sum_{\tau=n, p} {\nabla^m}{\rho_{\tau}}{\nabla^{k-m}}{\rho_{\tau}} \bigg] \Bigg\},
\end{aligned}
\end{equation}
\end{widetext}
where $\delta_{k4}$ denotes the Kronecker delta;
$f_{\tau}\left(\vec{r}, \vec{p}\right)$ is the nucleon
phase distribution function;
$C^{[n]}$, $D^{[n]}$, $E^{[n]}$ and $F^{[n]}$ are the combinations of Skyrme parameters $t_1^{[n]}$, $x_1^{[n]}$, $t_2^{[n]}$, $x_2^{[n]}$, i.e.
\begin{eqnarray}
\label{eq:defineCDEF}
    C^{\left[ n \right]}  &=& t_{1}^{\left[ n \right]} \left( 2 + x_{1}^{\left[ n \right]} \right) + t_{2}^{\left[ n \right]} \left( 2 + x_{2}^{\left[ n \right]} \right), \notag\\
    D^{\left[ n \right]} &=& -t_{1}^{\left[ n \right]} \left( 2 x_{1}^{\left[ n \right]} +1 \right) + t_{2}^{\left[ n \right]} \left( 2 x_{2}^{\left[ n \right]} +1 \right), \notag\\
    E^{\left[ n \right]}& =& \frac{i^{n}}{2^{n}} \left[
    t_{1}^{\left[ n \right]} \left(2 + x_{1}^{\left[ n \right]} \right) - t_{2}^{\left[ n \right]} \left( 2 + x_{2}^{\left[ n \right]} \right)
    \right], \notag\\
    F^{\left[ n \right]} &=& - \frac{i^{n}}{2^{n}} \left[
    t_{1}^{\left[ n \right]} \left( 2 x_{1}^{\left[ n \right]} +1 \right) + t_{2}^{\left[ n \right]} \left( 2 x_{2}^{\left[ n \right]} +1 \right)
    \right],
\end{eqnarray}
with $n=2,4,6$.
Note that the number of parameters remains unchanged at twenty after the recombination.
The single-nucleon potential can then be calculated as $ U_{\tau} \left(\vec{r}, \vec{p}\right)=\delta V(\rho_n,\rho_p)/\delta \rho_\tau\left( \vec{r}, \vec{p} \right)$~($\tau = n$ or $p$).

In order to extend the newly developed N3LO Skyrme pseudopotential~\citep{Wang:2018yce,Wang:2023zcj}, which incorporates local, momentum-dependent, gradient and density-dependent terms, to the case of the octet baryons, it is convenient to
rewrite the potential energy density in Eq.~(\ref{eq:VNN_old}) in terms of explicit isospin index, in analogy to that of the momentum-dependent interaction~(MDI) model~\citep{Das:2002fr,Chen:2004si,Xu:2010re}, namely,
%----------------------------------------------------------------------------------------------------
\begin{widetext}
\begin{equation}
\label{eq:VNN_new}
\begin{aligned}
    V(\rho_n,\rho_p)=&
    A_{u}\rho_{n}\rho_{p}+\frac{A_{l}}{2}(\rho_{n}^{2}+\rho_{p}^{2})
    +\sum_{\alpha=1,3,5}{\rho^{\frac{\alpha}{3}}}\left[B^{[\alpha]}_{u}\rho_{n}\rho_{p}+\frac{B^{[\alpha]}_{l}}{2}(\rho_{n}^{2}+\rho_{p}^{2})\right]\\
    &+\sum_{k=2,4,6}\Bigg\{\iint d^{3}pd^{3}p^{\prime}{(\vec{p}-\vec{p}^{\,\prime})^{k}}
    \bigg[\frac{M^{[k]}_{u}}{\hbar^{k}} {f_{n}(\vec{r},\vec{p})f_{p}(\vec{r},{\vec{p}}^{\,\prime})}
    +\frac{M^{[k]}_{l}}{2\hbar^{k}}\Big({f_{n}(\vec{r},\vec{p})f_{n}(\vec{r},{\vec{p}}^{\,\prime})}+{f_{p}(\vec{r},\vec{p})f_{p}(\vec{r},{\vec{p}}^{\,\prime})}\Big)\bigg]\\
    &+\sum_{m=0}^{k}(-1)^{m}{\dbinom{k}{m}}\bigg[G^{[k]}_{u}{\nabla^m}{\rho_{n}}{\nabla^{k-m}}{\rho_{p}}
    +\frac{G^{[k]}_{l}}{2}\Big({\nabla^m}{\rho_{n}}{\nabla^{k-m}}{\rho_{n}}+{\nabla^m}{\rho_{p}}{\nabla^{k-m}}{\rho_{p}}\Big)\bigg]\Bigg\},
\end{aligned}
\end{equation}
%----------------------------------------------------------------------------------------------------
where $A_u=\frac{t_0(2+x_0)}{2}, A_l=\frac{t_0(1-x_0)}{2},
B^{[\alpha]}_u=\frac{{t^{[\alpha]}_3}(2+x^{[\alpha]}_3)}{12}, B^{[\alpha]}_l=\frac{{t^{[\alpha]}_3}(1-x^{[\alpha]}_3)}{12},
M^{[k]}_u=\frac{C^{[k]}}{8(1+\delta_{k4})},M^{[k]}_l=\frac{C^{[k]}+D^{[k]}}{8(1+\delta_{k4})},
G^{[k]}_u=\frac{E^{[k]}}{8(1+\delta_{k4})},G^{[k]}_l=\frac{E^{[k]}+F^{[k]}}{8(1+\delta_{k4})}$, with $\alpha = (1,3,5)$ and $k = (2,4,6)$.

The explicit isospin index in Eq.~(\ref{eq:VNN_new}) makes it easier to generalize the newly extended N3LO Skyrme pseudopotential to include the YN and YY interactions in terms of different baryon flavors.
In particular, to incorporate hyperons into the N3LO Skyrme pseudopotential,
we assume that the YN and YY interactions have similar density, momentum, and isospin
dependence as the NN interaction, and thus one can express the potential energy density contributions from any two types of octet baryons ($b$ and $b^{\prime}$) utilizing the following form:
%----------------------------------------------------------------------------------------------------
\begin{equation}
\label{Vbb1}
\begin{aligned}
    V_{b{b^{\prime}}}=&\sum_{{\tau_b},{\tau^{\prime}_{b^{\prime}}}}\Bigg\{\left(\frac{A_{b{b^{\prime}}}}{2}{\rho_{\tau_b}}{\rho_{\tau^{\prime}_{b^{\prime}}}}
    +\frac{A^{\prime}_{b{b^{\prime}}}}{2}{\tau_b}{\tau^{\prime}_{b^{\prime}}}{\rho_{\tau_b}}{\rho_{\tau^{\prime}_{b^{\prime}}}}\right)
    +\sum_{\alpha=1,3,5}{\rho^{\frac{\alpha}{3}}}\left(\frac{B^{[\alpha]}_{b{b^{\prime}}}}{2}{\rho_{\tau_b}}{\rho_{\tau^{\prime}_{b^{\prime}}}}
    +\frac{B^{\prime[\alpha]}_{b{b^{\prime}}}}{2}{\tau_b}{\tau^{\prime}_{b^{\prime}}}{\rho_{\tau_b}}{\rho_{\tau^{\prime}_{b^{\prime}}}}\right)\\
    &+\sum_{k=2,4,6}\bigg[
     \iint d^{3}pd^{3}p^{\prime}{(\vec{p}-\vec{p}^{\,\prime})^{k}}\Big({\frac{M^{[k]}_{bb^{\prime}}}{2\hbar^{k}}}{f_{\tau_b}(\vec{r},\vec{p}
    )f_{\tau^{\prime}_{b^{\prime}}}(\vec{r},{\vec{p}}^{\,\prime})}+{\frac{M^{\prime [k]}_{bb^{\prime}}}{2\hbar^{k}}}{\tau_b}{\tau^{\prime}_{b^{\prime}}}{f_{\tau_b}(\vec{r},\vec{p}
    )f_{\tau^{\prime}_{b^{\prime}}}(\vec{r},{\vec{p}}^{\,\prime})}\Big)\\
    &+\sum_{m=0}^{k}(-1)^{m}{\dbinom{k}{m}}\Big(\frac{G^{[k]}_{bb^{\prime}}}{2}{\nabla^m}{\rho_{\tau_b}}{\nabla^{k-m}}{\rho_{\tau^{\prime}_{b^{\prime}}}}+\frac{G^{\prime [k]}_{bb^{\prime}}}{2}{\tau_b}{\tau^{\prime}_{b^{\prime}}}{\nabla^m}{\rho_{\tau_b}}{\nabla^{k-m}}{\rho_{\tau^{\prime}_{b^{\prime}}}}\Big)\bigg]\Bigg\},
\end{aligned}
\end{equation}
\end{widetext}
%----------------------------------------------------------------------------------------------------
where $b$ and $b^{\prime}$ represent the octet baryons ($N,\Lambda,\Sigma,\Xi$), and $\tau_b$ and $\tau_{b^{\prime}}$ denote the third component of isospin for the corresponding octet baryons $b$ and $b^{\prime}$, respectively.
Specifically, in this work, we assume $\tau_{N}=-1$ for neutrons and $1$ for protons; $\tau_{\Lambda}=0$ for $\Lambda$; $\tau_{\Sigma}=-1$, $0$ and $1$ for $\Sigma^{-}$, $\Sigma^{0}$ and $\Sigma^{+}$, respectively; $\tau_{\Xi}=-1$ for $\Xi^{-}$ and $1$ for $\Xi^{0}$.
For nucleonic system which only consists of neutrons and protons, the interaction parameters in Eq.~(\ref{Vbb1}) are given by
%----------------------------------------------------------------------------------------------------
\begin{equation}
\label{ABNN}
\begin{aligned}
    &A_{NN}=\frac{A_l+A_u}{2},~
    A^{\prime}_{NN}=\frac{A_l-A_u}{2},\\
    &X^{[\alpha]}_{NN}=\frac{X^{[\alpha]}_l+X^{[\alpha]}_u}{2},~
    X^{\prime[\alpha]}_{NN}=\frac{X^{[\alpha]}_l-X^{[\alpha]}_u}{2},
\end{aligned}
\end{equation}
%----------------------------------------------------------------------------------------------------
where we have $X={B}$ with $\alpha=(1,3,5)$, or $X=(M,G)$ with $\alpha=(2,4,6)$.

For octet baryons, we assume that the interaction parameters of YN and YY interactions are all proportional to their corresponding counterparts in nucleon-nucleon interaction, namely
%----------------------------------------------------------------------------------------------------
\begin{equation}
\label{AB}
\begin{aligned}
    &A_{b{b^{\prime}}}=f_{A_{b{b^{\prime}}}}A_{NN},~
    A^{\prime}_{b{b^{\prime}}}=f_{A^{\prime}_{b{b^{\prime}}}}A^{\prime}_{NN},\\
    &X^{[\alpha]}_{b{b^{\prime}}}=f_{X^{[\alpha]}_{b{b^{\prime}}}}X^{[\alpha]}_{NN},~
    X^{\prime[\alpha]}_{b{b^{\prime}}}=f_{X^{\prime[\alpha]}_{b{b^{\prime}}}}X^{\prime[\alpha]}_{NN},
    % &(X=\{B\},\alpha=\{1,3,5\};\\
    % &~X=\{C,D\},\alpha=\{1,2,3\})
\end{aligned}
\end{equation}
%----------------------------------------------------------------------------------------------------
and the meanings of $X$ and $\alpha$ in Eq.~(\ref{AB}) are the same as that in Eq.~(\ref{ABNN}).
The determination of the scaling parameters $f$, denoted by various subscripts in Eq.~(\ref{AB}), will be discussed subsequently.

The total potential energy density of hypernuclear system can then be obtained by summing over all types of $V_{b{b^{\prime}}}$ as
%----------------------------------------------------------------------------------------------------
\begin{equation}
\label{V:HP}
\begin{aligned}
    V_{HP}=&\frac{1}{2}\sum_{b,{b^{\prime}}}(1+{\delta}_{b{b^{\prime}}})V_{b{b^{\prime}}}.
\end{aligned}
\end{equation}
%----------------------------------------------------------------------------------------------------
By performing variation of $V_{HP}$, one can obtain the single-particle potential of the baryon species of $\tau_b$, i.e.,
%----------------------------------------------------------------------------------------------------
\begin{widetext}
\begin{equation}
\label{U:tau}
\begin{aligned}
    U_{\tau_b}(p)=&\frac{{\delta}V_{HP}}{{\delta}{\rho_{\tau_b}}}\\
    =&\sum_{b^{\prime}}\sum_{\tau^{\prime}_{b^{\prime}}}(1+{\delta_{b{b^{\prime}}}})
    \Bigg\{\left(\frac{A_{b{b^{\prime}}}}{2}{\rho_{\tau^{\prime}_{b^{\prime}}}}
    +\frac{A^{\prime}_{b{b^{\prime}}}}{2}{\tau_b}{\tau^{\prime}_{b^{\prime}}}{\rho_{\tau^{\prime}_{b^{\prime}}}}\right)
    +\sum_{\alpha=1,3,5}{\rho^{\frac{\alpha}{3}}}\left(\frac{B^{[\alpha]}_{b{b^{\prime}}}}{2}{\rho_{\tau^{\prime}_{b^{\prime}}}}
    +\frac{B^{\prime[\alpha]}_{b{b^{\prime}}}}{2}{\tau_b}{\tau^{\prime}_{b^{\prime}}}{\rho_{\tau^{\prime}_{b^{\prime}}}}\right)\\
    &+\sum_{k=2,4,6}\bigg[\int d^{3}p^{\prime}{(\vec{p}-\vec{p}^{\,\prime})^{k}}{\Big({\frac{M^{[k]}_{bb^{\prime}}}{2\hbar^{k}}}{f_{\tau^{\prime}_{b^{\prime}}}(\vec{r},{\vec{p}}^{\,\prime})}+{\frac{M^{\prime [k]}_{bb^{\prime}}}{2\hbar^{k}}}{\tau_b}{\tau^{\prime}_{b^{\prime}}}{f_{\tau^{\prime}_{b^{\prime}}}(\vec{r},{\vec{p}}^{\,\prime})}\Big)}\\
    &+{\Big({\frac{G^{[k]}_{bb^{\prime}}}{2}}{2^{k}}{\nabla^{k}}{\rho_{\tau^{\prime}_{b^{\prime}}}}\Big)+\Big({\frac{G^{\prime [k]}_{bb^{\prime}}}{2}}{2^{k}}{\tau_b}{\tau^{\prime}_{b^{\prime}}}{\nabla^{k}}{\rho_{\tau^{\prime}_{b^{\prime}}}}\Big)}\bigg]\Bigg\}\\
    &+\frac{1}{2}\sum_{b^{\prime},b^{\prime\prime}}\sum_{{\tau}_{b^{\prime}},{\tau}^{\prime}_{b^{\prime\prime}}}(1+{\delta_{{b^{\prime}}{b^{\prime\prime}}}})\Bigg[\sum_{\alpha=1,3,5}{\frac{\alpha}{3}}{\rho^{\frac{\alpha}{3}-1}}\left(\frac{B^{[\alpha]}_{b^{\prime}{b^{\prime\prime}}}}{2}{\rho_{\tau_{b^{\prime}}}}{\rho_{\tau^{\prime}_{b^{\prime\prime}}}}
    +\frac{B^{\prime[\alpha]}_{b^{\prime}{b^{\prime\prime}}}}{2}{\tau_{b^{\prime}}}{\tau^{\prime}_{b^{\prime\prime}}}{\rho_{\tau_{b^{\prime}}}}{\rho_{\tau^{\prime}_{b^{\prime\prime}}}}\right)\Bigg].
\end{aligned}
\end{equation}
\end{widetext}
%----------------------------------------------------------------------------------------------------

%----------------------------------------------------------------------------------------------------
\subsection{Determination of parameters}
In the present work, we mainly focus on infinite uniform nuclear and hypernuclear matter for which the gradient terms vanish.
For nuclear matter,
by excluding the parameters associated with gradient terms (i.e., $E^{[2]}$, $E^{[4]}$, $E^{[6]}$, $F^{[2]}$, $F^{[4]}$, $F^{[6]}$, or equivalently $G_u^{[2]}$, $G_u^{[4]}$, $G_u^{[6]}$, $G_l^{[2]}$, $G_l^{[4]}$, $G_l^{[6]}$), the recently developed extended N3LO Skyrme pseudopotential for nucleons~\citep{Wang:2023zcj} can then be characterized solely by fourteen microscopic Skyrme parameters (or their combinations): $t_0$, $t_{3}^{[1]}$, $t_{3}^{[3]}$, $t_{3}^{[5]}$, $x_0$, $x_{3}^{[1]}$, $x_{3}^{[3]}$, $x_{3}^{[5]}$, $C^{[2]}$, $C^{[4]}$, $C^{[6]}$, $D^{[2]}$, $D^{[4]}$ and $D^{[6]}$,
or equivalently $A_u$, $A_l$, $B_u^{[1]}$, $B_u^{[3]}$, $B_u^{[5]}$, $B_l^{[1]}$, $B_l^{[3]}$, $B_l^{[5]}$, $M_u^{[2]}$, $M_u^{[4]}$, $M_u^{[6]}$, $M_l^{[2]}$, $M_l^{[4]}$, $M_l^{[6]}$.
These microscopic parameters can be determined by fourteen macroscopic quantities~\citep{Wang:2023zcj}: $\rho_0$, $E_{0}(\rho_0)$, $K_0$, $J_0$, $a_2$, $a_4$, $a_6$, $b_2$, $b_4$, $b_6$, $E_{\mathrm{sym}}(\rho_0)$, $L$, $K_{\mathrm{sym}}$ and $J_{\mathrm{sym}}$.
Here, $a_2$, $a_4$ and $a_6$ represent the polynomial momentum expansion coefficients of the single-nucleon potential (optical potential) in SNM at $\rho_0$, namely $U_{0}(\rho_0,p)$, while $b_2$, $b_4$ and $b_6$ correspond to those of the (first-order) nuclear symmetry potential at $\rho_0$, namely $U_{\mathrm{sym}}(\rho_0,p)$~\citep{Wang:2023zcj}:
\begin{eqnarray}
\label{eq:UU0}
%\begin{split}
    U_{0}(\rho_0,p)&=&U_{0}(\rho_0,0) + a_2 \left(\frac{p}{\hbar}\right)^2  + a_{4} \left(\frac{p}{\hbar}\right)^4 \notag \\
    &+& a_{6} \left(\frac{p}{\hbar}\right)^6,\\
%    \end{split}
%\end{eqnarray}
%\begin{eqnarray}
\label{eq:UUsym}
%\begin{split}
    U_{\mathrm{sym}}(\rho_0,p)&=&U_{\mathrm{sym}}(\rho_0,0) + b_{2} \left(\frac{p}{\hbar}\right)^2  + b_{4} \left(\frac{p}{\hbar}\right)^4  \notag \\
    &+& b_{6} \left(\frac{p}{\hbar}\right)^6.
%\end{split}
\end{eqnarray}
In the above expressions, $U_{0}(\rho_0,0)$, $a_2$, $a_4$ and $a_6$ are determined by fitting the nucleon optical potential $U_{\mathrm{opt}}$ obtained from high energy nucleon-nucleus elastic scattering data, while $U_{\mathrm{sym}}(\rho_0,0)$, $b_2$, $b_4$ and $b_6$ are obtained from fitting microscopic calculations, as detailed in Refs.~\citep{Wang:2018yce,Wang:2023zcj}. Following our previous work~\citep{Wang:2023zcj}, we fix the values of those parameters, i.e., $a_2=6.52\,\mathrm{MeV\,fm}^{2}$, $a_4=-0.126\,\mathrm{MeV\,fm}^{4}$, $a_6=8.13\times10^{-4}\,\mathrm{MeV\,fm}^{6}$, $b_2=-3.00\,\mathrm{MeV\,fm}^{2}$, $b_4=0.0780\,\mathrm{MeV\,fm}^{4}$ and $b_6=-7.00\times10^{-4}\,\mathrm{MeV\,fm}^{6}$. Note that both $U_{0}(\rho_0,0)$ and $U_{\mathrm{sym}}(\rho_0,0)$ are not independent variables and their values are determined by the input macroscopic quantities via the Hugenholtz-Van Hove (HVH) theorem~\citep{Hugenholtz:1958zz,SATPATHY199985,Xu:2010fh,Chen:2011ag,Cai:2012en}.

For the NN interaction parameters, apart from the six fixed polynomial momentum expansion coefficients of $U_{0}(\rho_0,p)$ and $U_{\rm sym}(\rho_0,p)$ mentioned above, namely, $a_2$, $a_4$, $a_6$, $b_2$, $b_4$, $b_6$, we have also chosen $\rho_0=0.16\, \mathrm{fm}^{-3}$, $E_0(\rho_0)=-16\,\rm{MeV}$, $K_0=230\,\rm{MeV}$, $J_0=-383\,\rm{MeV}$, $E_{\rm sym}(\rho_0)=32\,\rm{MeV}$ and $L=35\,\rm{MeV}$ as default values.
The default values of $\rho_0$, $E_0(\rho_0)$, $K_0$ and $J_0$ for SNM are the same as those used in our previous work~\citep{Wang:2023zcj}.
Furthermore, setting $K_{\rm sym}=-300\rm~MeV$ and $J_{\rm sym}=720\rm~MeV$, together with the default set of twelve macroscopic quantities: $a_2$, $a_4$, $a_6$, $b_2$, $b_4$, $b_6$, $\rho_0$, $E_0(\rho_0)$, $K_0$, $J_0$, $E_{\rm sym}(\rho_0)$ and $L$, leads to a standard interaction for nucleonic matter description, denoted as HSL35 in the present work.
As we will see in the following, the interaction HSL35 not only satisfies the constraints of the mass-radius relation (for both NSs and HSs) from astrophysical observations but also aligns with the microscopic calculations regarding EOS of PNM.
Further details will be discussed in the next section.

Once the interaction parameters of NN interaction are provided, one can determine the scaling parameters $f$ by fitting the single-particle potential $U^{(b^{\prime})}_b$, which represents the potential felt by a baryon $b$ in the matter of baryon ${b^{\prime}}$, to match the corresponding results obtained from experiments or microscopic calculations.
The scaling parameters
can be classified into various types according to whether they correspond to the local terms, the density-dependent terms or the momentum-dependent terms in the single-particle potential.
In addition, these types of parameters can further be categorized according to whether the corresponding terms are isospin-dependent or isospin-independent.
Although each parameter of the YN interaction has a unique corresponding scaling parameter $f$ that may in principle be different from others, in the present work, we usually assume that the scaling parameters $f$ are equal if they belong to the same class for the sake of simplicity.
For instance, $f_{B^{[1]}_{N\Sigma}},f_{B^{[3]}_{N\Sigma}}$ and $f_{B^{[5]}_{N\Sigma}}$ are all density-dependent but isospin-indenpendent scaling parameters, and for simplicity, we assume that they are equal to each other and they are treated as a single free parameter.

For the $\Lambda N$ interaction, there have relatively more reliable constraints on the potential of $\Lambda$ in SNM~\citep{Kochankovski:2023trc} compared to other hyperons. The isospin-dependent scaling parameters of $\Lambda$ with primes in their superscripts are meaningless, as they are always multiplied by $\tau_{\Lambda}=0$. The scaling parameter $f_{A_{N\Lambda}}=1.36$ is obtained from fitting the $\Lambda$ potential in SNM at saturation density and zero momentum~\citep{Millener:1988hp,Friedman:2024epf}:
%----------------------------------------------------------------------------------------------------
\begin{equation}
\label{ULambda}
\begin{aligned}
    U^{(N)}_{\Lambda}(\rho_N=\rho_0, p=0)=-28\,\rm MeV,
\end{aligned}
\end{equation}
%----------------------------------------------------------------------------------------------------
Furthermore, we obtain the scaling parameters $f_{B^{[1]}_{N\Lambda}} = f_{B^{[3]}_{N\Lambda}} = f_{B^{[5]}_{N\Lambda}} = 1.5$ by fitting the density dependence of the $\Lambda$ potential in SNM [i.e., $U^{(N)}_{\Lambda}(\rho_N, p=0)$] around $\rho_0$ predicted by $\chi \rm EFT$~\citep{Petschauer:2015nea}.
Similarly, to reproduce the momentum dependence of the $\Lambda$ potential in SNM at $\rho_0$ [i.e., $U^{(N)}_{\Lambda}(\rho_N=\rho_0, p)$] given by $\chi \rm EFT$, we fix $f_{M^{[2]}_{N\Lambda}} = f_{M^{[4]}_{N\Lambda}} = f_{M^{[6]}_{N\Lambda}} = 1.60$.

For the $\Sigma N$ interaction, to reproduce the density dependence of $\Sigma$ potential (at zero momentum) around $\rho_0$ in SNM predicted by $\chi \rm EFT$~\citep{Petschauer:2015nea}, we fix $f_{B^{[1]}_{N\Sigma}} = f_{B^{[3]}_{N\Sigma}} = f_{B^{[5]}_{N\Sigma}} = 1.30$.
Then we choose~\citep{Petschauer:2015nea}
%----------------------------------------------------------------------------------------------------
\begin{equation}
\label{USigma}
\begin{aligned}
    U^{(N)}_{\Sigma}(\rho_N=\rho_0, p=0)=11\,\rm MeV
\end{aligned}
\end{equation}
%----------------------------------------------------------------------------------------------------
in SNM and
%----------------------------------------------------------------------------------------------------
\begin{equation}
\label{USigma-}
\begin{aligned}
    U^{(N)}_{\Sigma^-}(\rho_N=\rho_0, p=0)=40\,\rm MeV
\end{aligned}
\end{equation}
%----------------------------------------------------------------------------------------------------
in PNM, which leads to $f_{A_{N\Sigma}}=0.88$ and $f_{A^{\prime}_{N\Sigma}}=f_{B^{\prime[1]}_{N\Sigma}}=f_{B^{\prime[3]}_{N\Sigma}}=f_{B^{\prime[5]}_{N\Sigma}}=2.41$, respectively.
Additionally, we set $f_{M^{[2]}_{N\Sigma}}=f_{M^{[4]}_{N\Sigma}}=f_{M^{[6]}_{N\Sigma}}=1.10$ and $f_{M^{\prime[2]}_{N\Sigma}}=f_{M^{\prime[4]}_{N\Sigma}}=f_{M^{\prime[6]}_{N\Sigma}}=3.70$, based on the momentum dependence of $\Sigma$ at $\rho_0$ in SNM and of $\Sigma^-$ at $\rho_0$ in PNM, calculated using $\chi \rm EFT$.

The procedure for determining the scaling parameters of the $\Xi N$ interaction is analogous to that for the $\Sigma N$ interaction.
It should be noted that while there exist calculations of the potentials for the $\Xi N$ interaction using LQCD~\citep{Inoue:2018axd}, no such calculations have been performed using $\chi \rm EFT$.
Consequently, we select the following LQCD results for the $\Xi N$ interaction~\citep{Inoue:2018axd}, i.e.,
%----------------------------------------------------------------------------------------------------
\begin{equation}
\label{UXi}
\begin{aligned}
    U^{(N)}_{\Xi}(\rho_N=\rho_0, p=0)=-4\,\rm MeV
\end{aligned}
\end{equation}
%----------------------------------------------------------------------------------------------------
in SNM and
%----------------------------------------------------------------------------------------------------
\begin{equation}
\label{UXi-}
\begin{aligned}
    U^{(N)}_{\Xi^-}(\rho_N=\rho_0, p=0)=7\,\rm MeV
\end{aligned}
\end{equation}
%----------------------------------------------------------------------------------------------------
in PNM, which respectively lead to $f_{A_{N\Xi}}=f_{B^{[1]}_{N\Xi}}=f_{B^{[3]}_{N\Xi}}=f_{B^{[5]}_{N\Xi}}=0.52$ and $f_{A^{\prime}_{N\Xi}}=f_{B^{\prime[1]}_{N\Xi}}=f_{B^{\prime[3]}_{N\Xi}}=f_{B^{\prime[5]}_{N\Xi}}=1.25$.
By setting $f_{M^{[2]}_{N\Xi}}=f_{M^{[4]}_{N\Xi}}=f_{M^{[6]}_{N\Xi}}=0.20$, we can reproduce the momentum dependence of $\Xi$ in SNM at $\rho_0$, as predicted by LQCD.
Similarly, when fixing $f_{M^{[2]\prime}_{N\Xi}}=f_{M^{\prime[4]}_{N\Xi}}=f_{M^{\prime[6]}_{N\Xi}}=0.50$, a relatively good fit is achieved for the momentum dependence of $\Xi^-$ in PNM at $\rho_0$, as obtained from LQCD.

The empirical ranges of $U^{(N)}_{Y}(\rho_N=\rho_0,p=0)$ for $\Sigma$ and $\Xi$ in SNM are estimated to approximately be $(10,50)$\,MeV and $(-24,-10)$\,MeV \citep{Friedman:2007zza,Friedman:2021rhu,Kochankovski:2023trc}, respectively.
Although the chosen value of $\Xi$ from LQCD falls slightly outside its empirical range, our aforementioned choices remain acceptable due to the large uncertainties regarding $\Xi$ properties. Nevertheless, we have checked that varying $U^{(N)}_{\Xi}(\rho_N=\rho_0, p=0)$ from $-4$ to $-16$~MeV (by factor four) only has small effects and does not change our conclusion.

Finally, there exists six types of YY interactions, which are poorly known for the moment.
To simplify the problem for a given YY interaction, we assume that the scaling parameters belonging to different classes are identical and only one adjustable scaling parameter exists.
The single scaling parameter can be determined using the equation concerning $U^{(Y^{\prime})}_{Y}(\rho_{Y^{\prime}}=\rho_0, p=0)$ in symmetric hypernuclear matter, similar to the approach employed for YN interaction.
Since the properties of the YY interaction are still ambiguous at present, following Refs.~\citep{Schaffner:1993qj,Xu:2010re}, we simply set
%----------------------------------------------------------------------------------------------------
\begin{equation}
\label{UYY}
\begin{aligned}
    U^{(Y^{\prime})}_{Y}(\rho_{Y^{\prime}}=\rho_0, p=0)=-40\,\rm MeV.
\end{aligned}
\end{equation}
%----------------------------------------------------------------------------------------------------
We note that varying the above value by factor two only has minor effects on our results.

%-------------------------------------------------------------------------------------------------------------------
\begin{figure*}[t!]
\centering
\includegraphics[width=1.98\columnwidth]{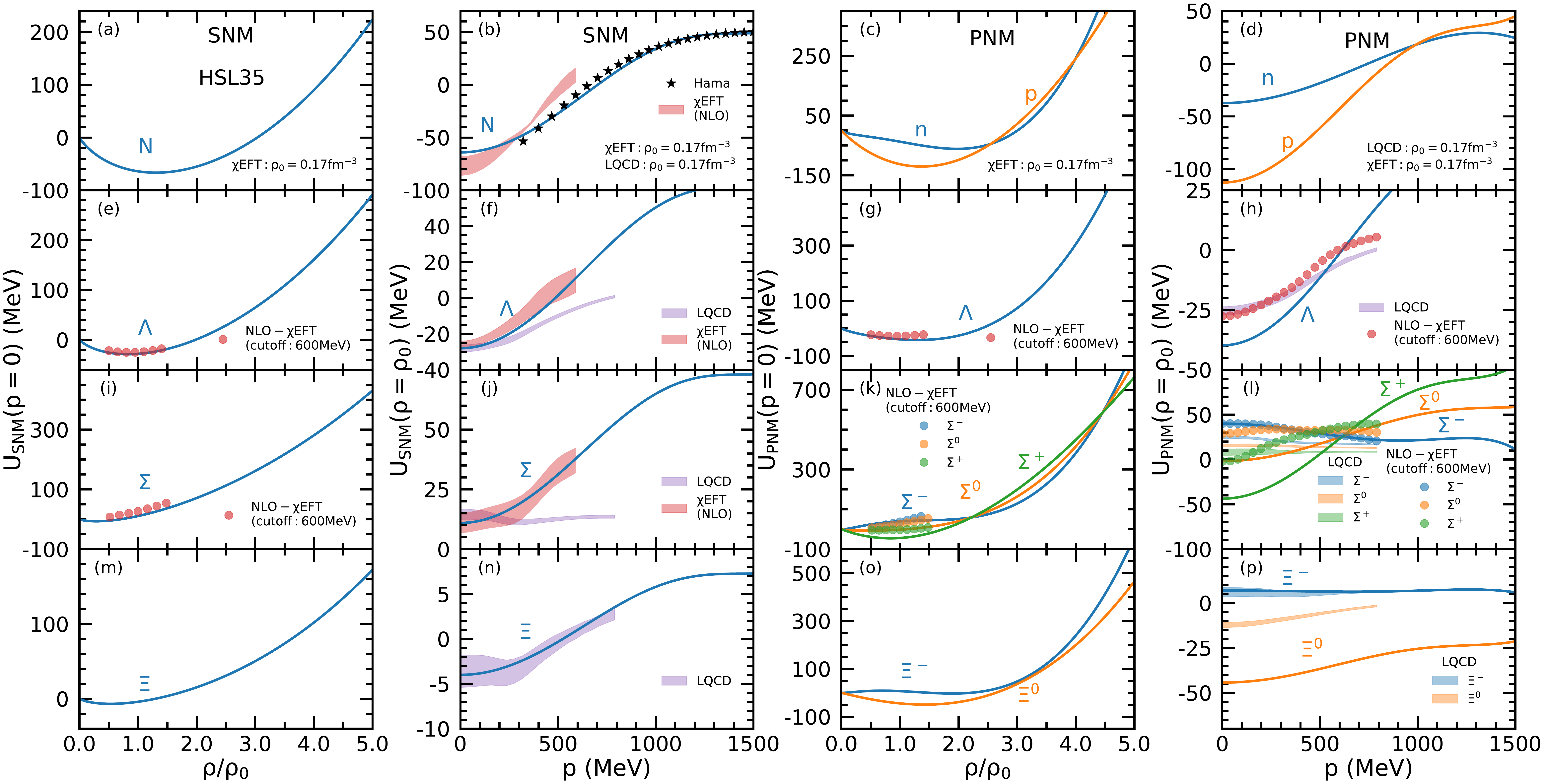}
\caption{The density (first and third columns) and momentum (second and fourth columns) dependence of single-particle potentials for octet baryons in SNM (left two columns) and PNM (right two columns) with the HSL35 interaction. The experimental nucleon optical potential~\citep{Hama:1990vr,Cooper:1993nx}, the calculations from $\chi \rm EFT$~\citep{Petschauer:2015nea} and LQCD~\citep{Inoue:2018axd} are also included for comparison.}
\label{fig:potential}
\end{figure*}
%-------------------------------------------------------------------------------------------------------------------

In Fig.~\ref{fig:potential}, we present the single-particle potentials of various octet baryons obtained using the HSL35 interaction, alongside the experimentally derived nucleon optical potential~\citep{Hama:1990vr,Cooper:1993nx}, the calculations based on chiral effective field theory~\citep{Petschauer:2015nea} and the results from LQCD~\citep{Inoue:2018axd}.
It is seen from Fig.~\ref{fig:potential} that the interaction HSL35 can reasonably describe the density and momentum dependence of the single-particle potentials for various octet baryons.
It should be pointed out that when we fix $U^{(N)}_{\Sigma}(\rho_N=\rho_0)$ in SNM as shown in Eq.\,(\ref{USigma}) and $U^{(N)}_{\Sigma^-}(\rho_N=\rho_0)$ in PNM as shown in Eq.\,(\ref{USigma-}), $U^{(N)}_{\Sigma^0}(\rho_N=\rho_0)$ and $U^{(N)}_{\Sigma^+}(\rho_N=\rho_0)$ have then also been fixed accordingly within the energy density functional used in our present work.
Similarly, $U^{(N)}_{\Xi^0}(\rho_N=\rho_0)$ is then determined accordingly once $U^{(N)}_{\Xi}(\rho_N=\rho_0)$ and $U^{(N)}_{\Xi^-}(\rho_N=\rho_0)$ are fixed as shown in Eq.\,(\ref{UXi}) and Eq.\,(\ref{UXi-}), respectively.
As a result, the momentum dependence of $\Sigma^-$ in PNM can be reproduced relatively more accurately than that of $\Sigma^0$ and $\Sigma^+$, and the momentum dependence of $\Xi^-$ is reproduced better than that of $\Xi^0$, as the potentials of $\Sigma^-$ and $\Xi^-$ in PNM have been used in our parameter calibration.
Nevertheless, because of the uncertainty associated with hyperonic interactions, the HSL35 interaction already provides a reasonable description of both experimental data and microscopic calculations overall.

%-------------------------------------------------------------------------------------------------------------------
\begin{table*}[t!]
\centering
\caption{Parameters for the NN interaction and the scaling parameters $f$ for the NN, YN and YY interactions in the HSL35 interaction.
Here, $f_{X_{NN}}$ and $f_{X_{YY}}$ represent the scaling parameters for the NN and YY interaction, respectively, and they are universal and independent of the different terms for NN and YY interactions.
The units of the parameters for NN interaction are as follows:
$A_{NN}$ and $A^{\prime}_{NN}$: $\mathrm{MeV}\, \mathrm{fm}^3$; $B^{[n]}_{NN}$ and $B^{\prime[n]}_{NN}$: $\mathrm{MeV}\, \mathrm{fm}^{n+3}$ ($n=1,3,5$); $M^{[n]}_{NN}$ and $M^{\prime[n]}_{NN}$: $\mathrm{MeV}\, \mathrm{fm}^{n+3}$ ($n=2,4,6$).
It is important to note that all scaling parameters denoted by $f$ are dimensionless.}

\footnotesize
\begin{tabular}{cccc|cccc|ccc|ccc}
\hline\hline
$A_{NN}$ & $B^{[1]}_{NN}$ & $B^{[3]}_{NN}$  & $B^{[5]}_{NN}$ & $A^{\prime}_{NN}$ & $B^{\prime[1]}_{NN}$ & $B^{\prime[3]}_{NN}$ & $B^{\prime[5]}_{NN}$ & $M^{[2]}_{NN}$ & $M^{[6]}_{NN}$ & $M^{[4]}_{NN}$ & $M^{\prime[2]}_{NN}$ & $M^{\prime[4]}_{NN}$ & $M^{\prime[6]}_{NN}$ \\
$-1380.34$ & $1626.28$ & $-504.55$ & $298.29$ & $52.13$ & $1439.44$ & $-4591.59$ & $3637.27$ & $43.62$ & $-0.83$ & $0.0051$ & $-21.86$ & $0.54$ & $-0.0044$\\
\hline
\multicolumn{1}{c|}{$f_{A_{N\Lambda}}$} & \multicolumn{3}{c|}{$f_{B^{[1]}_{N\Lambda}},f_{B^{[3]}_{N\Lambda}},f_{B^{[5]}_{N\Lambda}}$} & \multicolumn{4}{c|}{$f_{A^{\prime}_{N\Lambda}},f_{B^{\prime[1]}_{N\Lambda}},f_{B^{\prime[3]}_{N\Lambda}},f_{B^{\prime[5]}_{N\Lambda}}$} & \multicolumn{3}{c|}{$f_{M^{[2]}_{N\Lambda}},f_{M^{[4]}_{N\Lambda}},f_{M^{[6]}_{N\Lambda}}$} & \multicolumn{3}{c}{$f_{M^{\prime[2]}_{N\Lambda}},f_{M^{\prime[4]}_{N\Lambda}},f_{M^{\prime[6]}_{N\Lambda}}$}\\
\multicolumn{1}{c|}{1.36} & \multicolumn{3}{c|}{1.50} & \multicolumn{4}{c|}{-----} & \multicolumn{3}{c|}{1.60} & \multicolumn{3}{c}{-----}\\
\hline
\multicolumn{1}{c|}{$f_{A_{N\Sigma}}$} & \multicolumn{3}{c|}{$f_{B^{[1]}_{N\Sigma}},f_{B^{[3]}_{N\Sigma}},f_{B^{[5]}_{N\Sigma}}$}& \multicolumn{4}{c|}{$f_{A^{\prime}_{N\Sigma}},f_{B^{\prime[1]}_{N\Sigma}},f_{B^{\prime[3]}_{N\Sigma}},f_{B^{\prime[5]}_{N\Sigma}}$} & \multicolumn{3}{c|}{$f_{M^{[2]}_{N\Sigma}},f_{M^{[4]}_{N\Sigma}},f_{M^{[6]}_{N\Sigma}}$} & \multicolumn{3}{c}{$f_{M^{\prime[2]}_{N\Sigma}},f_{M^{\prime[4]}_{N\Sigma}},f_{M^{\prime[6]}_{N\Sigma}}$} \\
\multicolumn{1}{c|}{0.88} & \multicolumn{3}{c|}{1.30} & \multicolumn{4}{c|}{2.41} & \multicolumn{3}{c|}{1.10} & \multicolumn{3}{c}{3.70} \\
\hline
\multicolumn{4}{c|}{$f_{A_{N\Xi}},f_{B^{[1]}_{N\Xi}},f_{B^{[3]}_{N\Xi}},f_{B^{[5]}_{N\Xi}}$} & \multicolumn{4}{c|}{$f_{A^{\prime}_{N\Xi}},f_{B^{\prime[1]}_{N\Xi}},f_{B^{\prime[3]}_{N\Xi}},f_{B^{\prime[5]}_{N\Xi}}$} & \multicolumn{3}{c|}{$f_{M^{[2]}_{N\Xi}},f_{M^{[4]}_{N\Xi}},f_{M^{[6]}_{N\Xi}}$} & \multicolumn{3}{c}{$f_{M^{\prime[2]}_{N\Xi}},f_{M^{\prime[4]}_{N\Xi}},f_{M^{\prime[6]}_{N\Xi}}$} \\
\multicolumn{4}{c|}{0.52} & \multicolumn{4}{c|}{1.25} & \multicolumn{3}{c|}{0.20} & \multicolumn{3}{c}{0.50} \\
\hline
\multicolumn{2}{c}{$f_{X_{NN}}$} & \multicolumn{2}{c}{$f_{X_{ \Lambda\Lambda}}$} & \multicolumn{2}{c}{$f_{X_{\Sigma\Sigma}}$} & \multicolumn{2}{c}{$f_{X_{\Xi\Xi}}$} & \multicolumn{2}{c}{$f_{X_{\Sigma\Lambda}}$} & \multicolumn{2}{c}{$f_{X_{\Xi\Lambda}}$} & \multicolumn{2}{c}{$f_{X_{\Xi\Sigma}}$}\\
\multicolumn{2}{c}{1.00} & \multicolumn{2}{c}{0.67} & \multicolumn{2}{c}{0.61} & \multicolumn{2}{c}{0.62} & \multicolumn{2}{c}{1.33} & \multicolumn{2}{c}{1.33} & \multicolumn{2}{c}{1.22}\\
\hline
\hline
\end{tabular}
\label{tab1}
\end{table*}
%-------------------------------------------------------------------------------------------------------------------

In Table~\ref{tab1}, we list the parameters for the NN interaction in Eq.~(\ref{Vbb1}), as well as the scaling parameters $f$ for the YN and YY interactions.
These parameters are all derived from the standard interaction HSL35.
It is worth noting that
while varying the density dependence of the symmetry energy via taking various values of $K_{\rm sym}$ and $J_{\rm sym}$ (also $E_{\rm sym}(\rho_0)$ and $L$) may impact the parameters of the NN interaction, it does not change the scaling parameters $f$ whose values are assumed to be fixed at the corresponding ones in the HSL35 interaction.

%----------------------------------------------------------------------------------------------------
\subsection{Beta-stable hypernuclear matter}
The stable hyperon star matter satisfies the conditions of baryon number conservation, charge neutrality, and $\beta$-equilibrium:
\begin{equation}
\label{equilibrium}
\begin{aligned}
    \sum_i \rho_i b_i=& \rho_B,\\
    \sum_i \rho_i q_i =& 0,\\
    \mu_i =&\mu_b b_i-\mu_q q_i.
\end{aligned}
\end{equation}
In the above expressions, the subscript $i$ denotes the particle species, including nucleons, hyperons and leptons.
In this study, we assume that the hyperon star matter is neutrino free and the lepton category comprises electrons and muons.
The $b_i$, $q_i$ and $\mu_i$ represent the baryon number, (electric) charge number and chemical potential of the corresponding particles.
Due to the above equilibrium conditions, only two independent chemical potentials exist, i.e., the baryon chemical potential $\mu_b$ and the charge chemical potential $\mu_q$.
In addition, the $\rho_B$ corresponds to the total baryon number density.

For interacting octet baryon matter at zero temperature, the chemical potential of species $\tau_b$ can be expressed as follows:
\begin{equation}
\mu_{\tau_b} (p_{F{\tau_b}}) =m_{\tau_b}+
\frac{p_{F{\tau_b}}^2}{2m_{\tau_b}} + U_{\tau_b}(p_{F{\tau_b}}),
\end{equation}
where $m_{\tau_b}$ and $p_{F{\tau_b}}$ represent the baryon's rest mass and Fermi momentum, respectively.
The leptons are considered as relativistic free Fermi gas, and their chemical potentials are given by
\begin{equation}
    \mu_l(p_{Fl})=\sqrt{m_l^2+p_{Fl}^2}.
\end{equation}
The Fermi momentum of species $i$ can be determined using the formula $p_{Fi}=(3\pi^2 \rho_i)^{1/3}$.
By substituting the chemical potential formulae into Eq.~(\ref{equilibrium}) and solving for different total baryon densities $\rho_B$, one can then obtain the particle fractions of species $i$, denoted as $\rho_i/\rho_B$.
The total energy density $\epsilon$ consists of two components:
\begin{equation}
\begin{aligned}
    \epsilon &=\epsilon_H + \epsilon_L,
\end{aligned}
\end{equation}
where $\epsilon_H$ and $\epsilon_L$ denote the energy density of baryons and leptons, respectively.
Except for the total potential energy density of baryons $V_{HP}$ given in Eq.~(\ref{V:HP}), the mass term $\epsilon_{HM}$ and kinetic energy term $\epsilon_{HK}$ also contribute to $\epsilon_H$, which are computed by
\begin{equation}
\begin{aligned}
    \epsilon_H &=\epsilon_{HM} + \epsilon_{HK}+V_{HP},\\
    \epsilon_{HK} &=\sum_b \sum_{\tau_b}\frac{2}{(2\pi)^3}\int^{p_{F\tau_b}}_{0} (\frac{p^2}{2m_{\tau_b}})(4\pi p^2)dp\\
    &= \sum_b \sum_{\tau_b} \frac{p_{F\tau_b}^5}{10\pi^2m_{\tau_b}},\\
    \epsilon_{HM} &= \sum_b \sum_{\tau_b} \rho_{\tau_b} m_{\tau_b}.
\end{aligned}
\end{equation}
Since the electron's mass is very small, we treat the electron as massless particle, and the energy for a free electron can be approximated as $\sqrt{{m_e}^2+{p_e}^2} \approx p_e$.
Consequently, the energy density of leptons can be expressed as:
\begin{equation}
\begin{aligned}
    \epsilon_L =&\epsilon_e + \epsilon_\mu,\\
    \epsilon_e =&\frac{2}{(2\pi)^3}\int^{p_{Fe}}_{0} (p)(4\pi p^2)dp
    =\frac{p^4_{Fe}}{4 \pi^2},\\
    \epsilon_\mu =&\frac{2}{(2\pi)^3}\int^{p_{F\mu}}_{0} (\sqrt{m_{\mu}^2+p^2})(4\pi p^2)dp\\
    =& \frac{1}{4 \pi^2 } \left[p^{}_{F\mu} \mu^3_\mu -
    \frac{1}{2} m^2_\mu p^{}_{F\mu} \mu_\mu\right.\\
    &-\left.\frac{1}{2} m^4_\mu \ln\left(\frac{p^{}_{F\mu}+\mu_\mu}{m_\mu}\right) \right].\\
\end{aligned}
\end{equation}
According to the fundamental thermodynamic relation, the total pressure of $\beta$-stable and charge-neutral hypernuclear matter can be derived as follows:
\begin{equation}
    P=\left(\sum_b \sum_{\tau_b} \mu_{\tau_b} \rho_{\tau_b}+\sum_l \mu_l \rho_l\right)-\left(\epsilon_H+\epsilon_L\right).
\end{equation}
The thermodynamic consistency condition $P= \rho_B^2 \frac{d(\epsilon/\rho_B)}{d\rho_B}$ can also be verified using the aforementioned formulae.

To calculate the properties of NSs and HSs, one also needs the explicit information on the crust EOS.
In the present work, the core-crust transition density $\rho_t$, which separates the liquid core from the nonuniform inner crust, is determined self-consistently via the so-called dynamical method of Ref.~\citep{Xu:2009vi}.
In addition, the critical density between the inner and outer crusts is taken to be $\rho_{\rm{out}}=2.46\times10^{-4}\,\mathrm{fm}^{-3}$~\citep{Carriere:2002bx,Xu:2009vi,Xu:2008vz}.
For the outer crust with $\rho<\rho_{\mathrm{out}}$, we adopt the famous BPS (FMT) EOS~\citep{Baym:1971pw,Iida:1996hh}; for the inner crust with $\rho_{\mathrm{out}}<\rho<\rho_t$, where the so-called nuclear pasta could appear, we assume a parameterized EOS with the polytropic form $P=a+b \epsilon^{4/3}$ with $a$ and $b$ determined from the EOS at $\rho_{\rm{out}}$ and $\rho_t$~\citep{Carriere:2002bx,Xu:2009vi,Xu:2008vz}.

%-------------------------------------------------------------------------------------------------------------------
\section{Results and discussions}
\label{Sec:Result}

\subsection{Features of the HSL35 interaction}
As discussed previously, for infinite uniform nucleonic matter system, the N3LO Skyrme pseudopotential EDF has totally fourteen parameters, i.e., $t_0$, $t_{3}^{[1]}$, $t_{3}^{[3]}$, $t_{3}^{[5]}$, $x_0$, $x_{3}^{[1]}$, $x_{3}^{[3]}$, $x_{3}^{[5]}$, $C^{[2]}$, $C^{[4]}$, $C^{[6]}$, $D^{[2]}$, $D^{[4]}$ and $D^{[6]}$ (or equivalently $A_u$, $A_l$, $B_u^{[1]}$, $B_u^{[3]}$, $B_u^{[5]}$, $B_l^{[1]}$, $B_l^{[3]}$, $B_l^{[5]}$, $M_u^{[2]}$, $M_u^{[4]}$, $M_u^{[6]}$, $M_l^{[2]}$, $M_l^{[4]}$, $M_l^{[6]}$), which can be uniquely determined by fourteen macroscopic quantities~\citep{Wang:2023zcj}: $\rho_0$, $E_{0}(\rho_0)$, $K_0$, $J_0$, $a_2$, $a_4$, $a_6$, $b_2$, $b_4$, $b_6$, $E_{\mathrm{sym}}(\rho_0)$, $L$, $K_{\mathrm{sym}}$ and $J_{\mathrm{sym}}$.
In this work, our main motivation is to investigate how the high density symmetry energy influences the properties of HSs.
To this end,
we primarily explore the impact of the higher order symmetry energy parameters $K_{\rm sym}$ and $J_{\rm sym}$, which are used to feature the high density behavior of the symmetry energy,
while keep the remaining twelve macroscopic quantities [i.e., $\rho_0$, $E_{0}(\rho_0)$, $K_0$, $J_0$, $a_2$, $a_4$, $a_6$, $b_2$, $b_4$, $b_6$, $E_{\mathrm{sym}}(\rho_0)$ and $L$] fixed at their respective values in the standard interaction HSL35.

Before presenting the detailed results on NSs and HSs, we would like to mention some features about the EOS of SNM as well as the symmetry energy for the standard interaction HSL35.
Firstly,
for the properties of SNM,
the HSL35 prediction on pressure as a function of density up to about $5\rho_0$ is strictly in agreement with the constraints extracted from flow data in high-energy HICs~\citep{Danielewicz:2002pu}, as depicted in Fig.~1 of our previous work~\citep{Wang:2023zcj}.
Secondly,
for the magnitude of the symmetry energy at nuclear saturation density,
we take its value to be $E_{\rm sym}(\rho_0) = 32$~MeV for HSL35, since this is a widely accepted value for $E_{\rm sym}(\rho_0)$.
Thirdly,
for the slope parameter $L$, it is shown~\citep{Wang:2023zcj} that a value of $L$ significantly larger than $35$~MeV will lead to an inconsistency with the mass-radius relation from the observation of HESS J1731-347~\citep{Doroshenko:2022nwp}.
Conversely, if $L$ falls far below $35$~MeV, it becomes impossible to simultaneously satisfy both constraints imposed on the mass-radius relation for PSR J0030+0451 \citep{Miller:2019cac,Riley:2019yda} and PSR J0740+6620 \citep{Miller:2021qha,Riley:2021pdl} along with EOS of PNM derived through microscopic calculations~\citep{Huth:2020ozf,Zhang:2022bni}.
We would like to point out that $L = 35$~MeV in HSL35 is a quite acceptable $L$ value and it is safely within the one-sigma bound of the world-averaged value of $58.7 \pm 28$~MeV obtained by~\citep{Oertel:2016bki}, which perhaps represents so far the most complete survey about the $L$ constraints via adopting $53$ analyses from nuclear experiments and nuclear theory as well as neutron star observations, compared to the earlier survey by~\citep{Lattimer:2012xj} as well as by~\citep{Li:2013ola}, or the recent one by~\citep{Li:2021thg} which essentially only includes the constraints from astrophysical observations.

In addition,
for the $K_{\rm sym}$ parameter,
while several analyses have suggested $K_{\rm sym} = -100\pm 100$~MeV~\citep{Mondal:2017hnh,Margueron:2017lup,Grams:2022bbq,Li:2021thg}, a larger
range is still often used in the literature as usually the constraints depend strongly on the observables, the model and analysis methods~\citep{Li:2021thg,Zhang:2022sep}.
As pointed out by~\citep{Raithel:2019ejc} as well as by~\citep{Biswas:2020puz}, the value of $K_{\rm sym}$ can be as small as about $-330$~MeV in their analyses on
astrophysical observations, where prior correlations amongst certain combinations of nuclear parameters are not assumed and they augured correlations amongst certain combinations of nuclear parameters, assumed in most analyses, may lead to large $K_{\rm sym}$. Also the systematics of the density dependence of the symmetry energy based on a large samples of theoretical models suggests a lower limit of $-330$~MeV for $K_{\rm sym}$~\citep{Chen:2015gba}. So the choice for $K_{\rm sym} = -300$~MeV in HSL35 is reasonable.
Moreover, previous constraints summarized in Ref.~\citep{Li:2021thg} did not consider hyperon degrees of freedom in neutron stars, and our present work suggests a quite small value of $K_{\rm sym}$ is necessary when hyperons are considered in neutron stars, especially when the mass-radius relation from the observation of HESS J1731-347 is included in the analysis.
For the higher-order parameter $J_{\rm sym}$,
its value is still poorly known and there are
only some theoretical predictions with its value from about $-600$ to $+1000$~MeV~\citep{Chen:2009wv,Chen:2011ib,Dutra:2012mb,Dutra:2014qga,Tews:2016jhi}. Therefore, $J_{\rm sym} = 720$~MeV in HSL35 is also safely acceptable.

Furthermore, we note that
the $E_{\rm sym}(2\rho_0)$ is about $32$~MeV in HSL35, and this value is outside the interval $E_{\rm sym}(2\rho_0) = 51\pm 13$ MeV at a $68\%$ confidence level obtained by~\citep{Li:2021thg}, which is extracted from nine analyses of neutron star observables. Again, we would like to point out that the interval $E_{\rm sym}(2\rho_0) = 51\pm 13$ MeV depends on the data, model and method, e.g., the analyses summarized in~\citep{Li:2021thg} did not consider the small mass and radius for the central compact object of HESS J1731-347 as well as the hyperon degrees of freedom in neutron stars, which may have important influence on the extraction of $E_{\rm sym}(2\rho_0)$, as demonstrated in our present work. It should be mentioned that the current constraints of $E_{\rm sym}(2\rho_0)$ from terrestrial nuclear experiments can be as small as $30$~MeV within one-sigma uncertainty, see, e.g., Fig. 2.4 of~\citep{Sorensen:2023zkk}.
In short, the symmetry energy predicted by HSL35 is safely acceptable based on our
current knowledge of nuclear theory, nuclear experiments and astrophysical observations.

\begin{figure*}[t!]
\centering
\includegraphics[width=1.8\columnwidth]{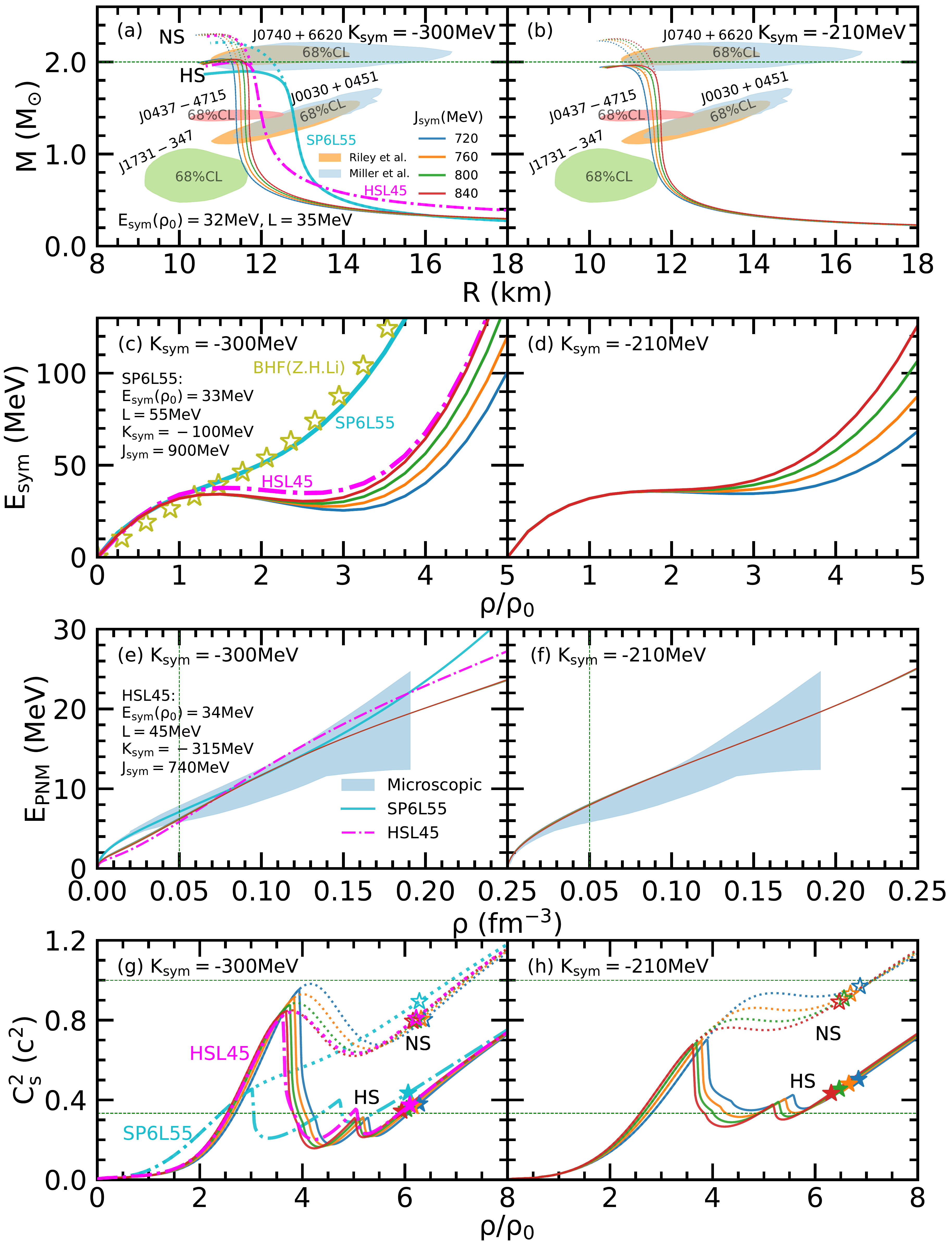}
\caption{The mass-radius relation for static HSs and NSs (first row), density dependence of the symmetry energy $E_{\rm sym}(\rho)$ (second row),
the EOS of PNM $E_{\rm PNM}(\rho)$ (third row) and density dependence of the squared sound speed $C^2_s(\rho/\rho_0)$ (fourth row) by varying individually $K_{\rm sym}$ and $J_{\rm sym}$ in the HSL35 interaction. The corresponding results with SP6L55 and HSL45 are included in the first column. The symmetry energy from the microscopic BHF calculation, denoted by BHF(Z.H.Li), is shown in panel~(c). The stars in the lowest row indicate the corresponding densities at the center of NSs and HSs with maximum mass configuration. For comparison, some typical constraints from astrophysical observations (i.e., the mass-radius of NSs at $68\%$ C.L.) and microscopic calculations are included in the first and third rows, respectively. Note in panels (e) and (f), the curves with different $J_{\rm sym}$ overlap each other. Further details can be found in the text.}
\label{fig:constraints}
\end{figure*}
%-------------------------------------------------------------------------------------------------------------------

\subsection{Effects of high-density symmetry energy}
By varying individually $K_{\rm sym}$ and $J_{\rm sym}$ in HSL35 with $K_{\rm sym}=-300$ and $-210$~MeV whereas $J_{\rm sym}$ spanning from $720$ to $840$~MeV in an interval of $40$~MeV, we show in Fig.~\ref{fig:constraints} the mass-radius relation for HSs and NSs [in panels (a) and (b)], the density dependence of the symmetry energy [in panels (c) and (d)], the EOS of PNM [in panels (e) and (f)] and the squared sound speed as a function of density for HS and NS matter [in panels (g) and(h)].
Also included in panels (a) and (b) of Fig.~\ref{fig:constraints}
are the constraints of the mass-radius relation obtained from astrophysical observations on PSR J0030+0451 ($M \approx 1.4M_\odot$) \citep{Miller:2019cac,Riley:2019yda}, PSR J0740+6620 ($M \approx 2M_\odot$) \citep{Miller:2021qha,Riley:2021pdl} and the recently reported PSR J0437-4715 ($M \approx 1.4M_\odot$) \citep{Choudhury:2024xbk} by NICER, as well as on the central compact object with a mass around $0.77M_\odot$ in the supernova remnant HESS J1731-347~\citep{Doroshenko:2022nwp}, at a confidence level (C.L.) of 68\%.
In addition, we also present,
in Fig.~\ref{fig:constraints}~(e) and (f), the combined constraint on $E_{\rm PNM}(\rho)$ incorporating various microscopic calculations~\citep{Huth:2020ozf,Zhang:2022bni}.

The case with $K_{\rm sym} = -300$~MeV and $J_{\rm sym} = 720$~MeV shown in the left column of Fig.~\ref{fig:constraints} [i.e., panels (a), (c), (e) and (g)] corresponds to the HSL35 interaction.
It is seen that
the HSL35 interaction predicts $M_{\rm TOV} = 2.03~(2.30)M_{\odot}$ for HSs~(NSs) and at the same time it can nicely describe the mass-radius relations obtained from astrophysical observations, especially the unusually low mass and small radius of HESS J1731-347.
Furthermore, the HSL35 interaction is also compatible with
the microscopic calculation constraint on the $E_{\rm PNM}(\rho)$ at density region of $\rho \gtrsim 0.05$~fm$^{-3}$.
Note that at lower densities (e.g., below about $0.05$~fm$^{-3}$), the pairing effects, which are not
considered in the mean-field calculations for $E_{\rm PNM}(\rho)$, may become considerable~\citep{Zhang:2019hlm}.
Moreover, one sees
although the squared sound speed ($C_s^2$) of NS matter exceeds the squared vacuum light speed ($c^2$) at very high densities, the $C_s^2$ at densities below the central density of the NS with $M_{\rm TOV}$ remains smaller than $c^2$, and so the causality condition is satisfied, guaranteeing that our results remain physically valid. Therefore, the HSL35 interaction provides a solution to the hyperon puzzle.

Now let's see how the higher order symmetry energy parameters $K_{\rm sym}$ and $J_{\rm sym}$ influence our results.
From the first row of Fig.~\ref{fig:constraints} [i.e., in panels (a) and (b)], one can see that the $M_{\rm TOV}$ for both NSs and HSs decreases with the increment of the $K_{\rm sym}$. In particular, the $M_{\rm TOV}$ of HSs (NSs) are $2.03~(2.30)M_{\odot}$ and $1.96~(2.23)M_{\odot}$ for $K_{\rm sym} = -300$ and $-210$~MeV, respectively, for a fixed value of $J_{\rm sym} = 720$~MeV.
We find that the $M_{\rm TOV}$ of HSs becomes smaller than $2M_{\odot}$ when $K_{\rm sym}$ is larger than about $-240$~MeV.
In addition, we note a large $K_{\rm sym}$ (e.g., $K_{\rm sym} > -210$~MeV) will violate the constraint on $E_{\rm PNM}(\rho)$ from microscopic calculations~\citep{Huth:2020ozf,Zhang:2022bni}.
As to the effects of $J_{\rm sym}$,
while the $J_{\rm sym}$ has obvious effects on the radius of NSs and HSs with a larger $J_{\rm sym}$ leading to a larger radius, it only has minor effects on the $M_{\rm TOV}$ for both NSs and HSs. For example, the $M_{\rm TOV}$ of HSs (NSs) are $2.03~(2.30)M_{\odot}$ and $2.03~(2.30)M_{\odot}$ for $J_{\rm sym} = 720$ and $840$~MeV, respectively, for a fixed value of $K_{\rm sym} = -300$~MeV. We note that the observed small radius of HESS J1731-347 cannot be obtained when $J_{\rm sym}$ is larger than about $760$~MeV.
On the other hand, a small $J_{\rm sym}$ will predict a too small radius for PSR J0030+0451 ($M \approx 1.4M_\odot$) to conform to the measured value by NICER~\citep{Miller:2019cac}.
These analyses and discussions suggest that a symmetry energy
with $K_{\rm sym} \approx -300$~MeV and $J_{\rm sym} \approx 720$~MeV, together with $E_{\rm sym}(\rho_0) \approx 32$~MeV and $L \approx 35$~MeV, can support massive HSs with $M_{\rm TOV} \gtrsim 2M_{\odot}$ and simultaneously it is compatible with the astrophysical constraints on the mass-radius relations by NICER and
HESS J1731-347 as well as the microscopic calculations on $E_{\rm PNM}(\rho)$.

From the second row of Fig.~\ref{fig:constraints}
[i.e., panels (c) and (d)],
it is seen that the symmetry energy with $K_{\rm sym} \approx -300$~MeV and $J_{\rm sym} \approx 720$~MeV
(together with $E_{\rm sym}(\rho_0) \approx 32$ MeV and $L \approx 35$ MeV) displays a quite soft density behavior around $2-3\rho_0$ but very stiff one above $4\rho_0$. The fast stiffening of the symmetry energy at high density is necessary to support a massive NS with mass larger than $2M_\odot$, given that the symmetry energy is soft at intermediate densities.
It is interesting to see from the lowest row of Fig.~\ref{fig:constraints}
[i.e., panels (g) and (h)] that the softening of the symmetry energy around $2-3\rho_0$ with smaller values of $K_{\rm sym}$ and $J_{\rm sym}$ can push the critical density for hyperon appearance to a larger value and thus leads to a larger $M_{\rm TOV}$ for HSs.
In addition, the hyperon appearance softens significantly the EOS of HS matter and thus leads to a clear peak for the squared sound speed around $4\rho_0$, which provides a natural explanation for the similar peak structure for squared sound speed observed in
Bayesian model-agnostic analyses on the multi-messenger data of NSs together with state-of-the-art theoretical calculations [see, e.g., Refs.~\citep{Legred:2021hdx,Marczenko:2022jhl,Han:2022rug,Cao:2023rgh,Marczenko:2023txe,Annala:2023cwx,Pang:2023dqj,Tang:2024jvs}].
In particular, we note that the HSL35 predicts a peak value of $C_s^2 = 0.95 c^2$ at $3.96\rho_0$.
This squared sound speed peak structure from HSL35 is consistent with the Bayesian model-agnostic analyses within a certain confidence interval.
For example, in Ref.~\citep{Legred:2021hdx}, the $C_s^2$ is found to reach a maximum of $0.75^{+0.25}_{-0.24} c^2$ at $3.60^{+2.25}_{-1.89}\rho_0$ at $90\%$ credibility.

Our results indicate that the softening of the symmetry energy at intermediate densities (around $2-3\rho_0$) plays an important role in enhancing the critical density for hyperon appearance and increasing the $M_{\rm TOV}$ of HSs.
For comparison,
we include the corresponding results
in the first column of Fig.~\ref{fig:constraints} [i.e., panels (a), (c), (e) and (g)] from a conventional interaction SP6L55 constructed in our previous work~\citep{Wang:2023zcj} and a new interaction HSL45.
For these two interactions, the fourteen macroscopic quantities [i.e., $\rho_0$, $E_{0}(\rho_0)$, $K_0$, $J_0$, $a_2$, $a_4$, $a_6$, $b_2$, $b_4$, $b_6$, $E_{\mathrm{sym}}(\rho_0)$, $L$, $K_{\mathrm{sym}}$ and $J_{\mathrm{sym}}$] are the same as those of the standard interaction HSL35 except for
$E_{\rm sym}(\rho_0)=33$~MeV, $L=55$~MeV, $K_{\rm sym}=-100$~MeV and $J_{\rm sym}=900$~MeV for SP6L55 while $E_{\rm sym}(\rho_0)=34$~MeV, $L=45$~MeV, $K_{\rm sym}=-315$~MeV and $J_{\rm sym}=740$~MeV for HSL45. Therefore, the SP6L55 has a much stiffer symmetry energy around intermediate densities compared to the HSL35 [see Fig.~\ref{fig:constraints}~(c)].
Also included in Fig.~\ref{fig:constraints}~(c) is the symmetry energy
from the microscopic Brueckner-Hartree-Fock calculations using the Bonn B nucleon-nucleon potentials together with compatible
microscopic meson-exchange three-body forces, denoted as BHF(Z.H.Li)~\citep{Li:2008zzt}.
One sees that
the density behavior of the symmetry energy for SP6L55 is very similar to that of BHF(Z.H.Li).
The SP6L55 predicts qualitatively different symmetry energy from the HSL35, and it gives a $M_{\rm TOV} = 1.90(2.21)M_{\odot}$ for HSs (NSs) and a large radius of about $13$~km at $M \approx 0.6 \sim 1.8 M_\odot$, failing to conform to the observed maximum mass for pulsars (in the case of HSs) and the astrophysical constraints on the mass-radius relation.
Compared to HSL35, the SP6L55 predict a much smaller critical density, namely, $\sim 3\rho_0$ for hyperon appearance, as shown in Fig.~\ref{fig:constraints}~(g), which softens the EOS of HS matter and thus leads to a smaller $M_{\rm TOV}$ for HSs.

On the other hand, as shown in Fig.~\ref{fig:constraints}~(c), the HSL45 displays quite similar density dependence of the symmetry energy, i.e., a soft density behavior around $2-3\rho_0$ but very stiff one above $4\rho_0$, with the HSL35 interaction, except for a little larger $L$ value. The main difference between HSL35 and HSL45 is that the HSL45 predicts a larger radii for NSs and HSs, leading to an inconsistency with the mass-radius relation from the observation of HESS J1731-347~\citep{Doroshenko:2022nwp}, as seen in Fig.~\ref{fig:constraints}~(a).
We note that the HSL45 predicts $M_{\rm TOV} = 2.0 M_\odot$, and actually when the $L$ is larger than $45$~MeV, it is hard (via varying $K_{\rm sym}$ and $J_{\rm sym}$) to satisfy simultaneously the constraints of the microscopic calculation of pure neutron matter and $M_{\rm TOV} \geq 2.0 M_\odot$ for hyperon stars, within our present theoretical framework.

%-------------------------------------------------------------------------------------------------------------------
\begin{figure}[t!]
\centering
\includegraphics[width=0.95\columnwidth]{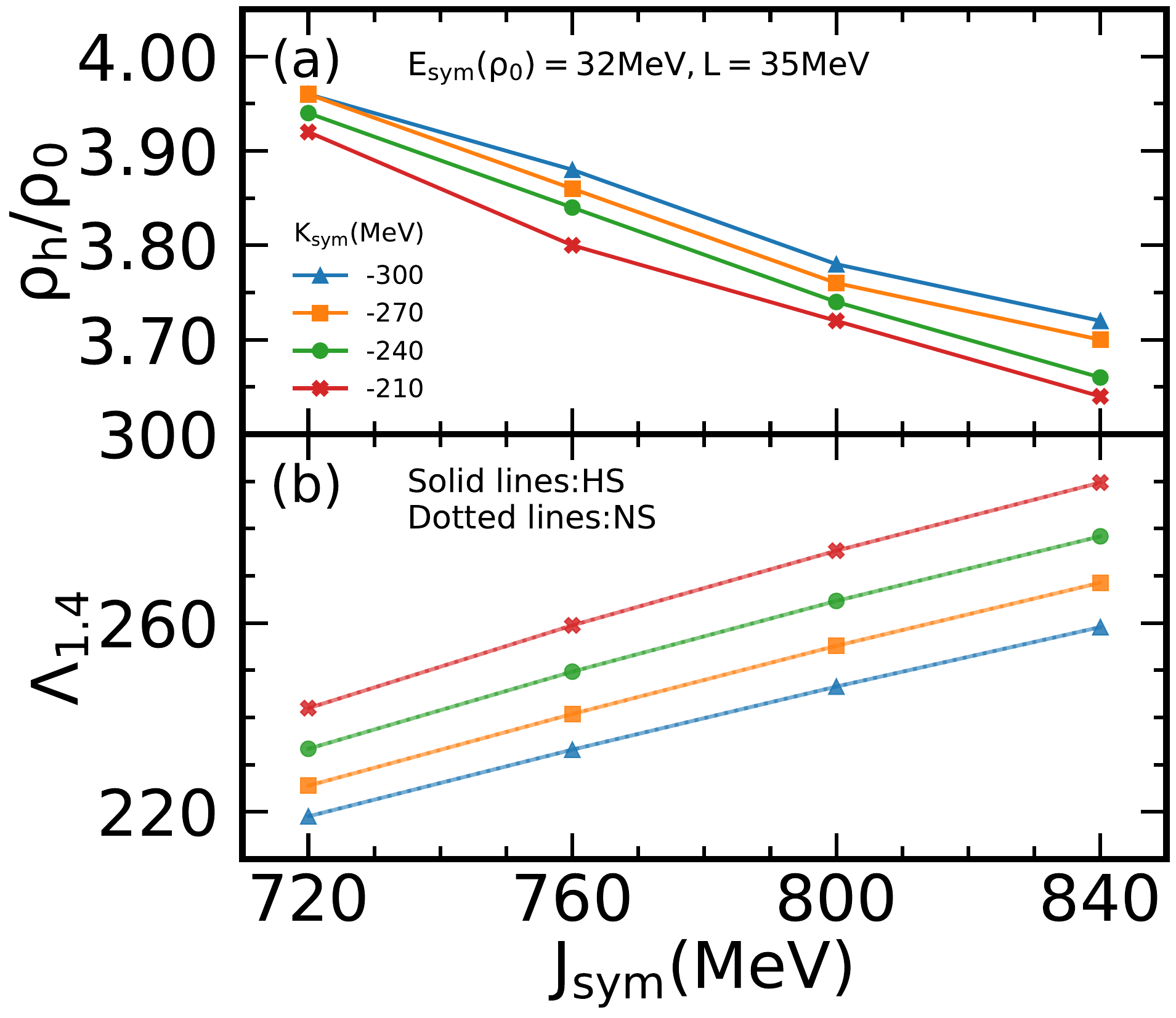}
\caption{The critical density $\rho_h$ for hyperon appearance in NS matter (a) and the tidal deformability $\Lambda_{1.4}$ of $1.4M_{\odot}$ NSs and HSs by varying individually $K_{\rm sym}$ and $J_{\rm sym}$ in the HSL35 interaction. Note the $\Lambda_{1.4}$ of NSs is the same as that of HSs and thus the symbols for NSs and HSs in panel (b) are exactly coincided. See the text for more details.}
\label{fig:lambda}
\end{figure}
%-------------------------------------------------------------------------------------------------------------------

In order to see more clearly the effects of $K_{\rm sym}$ and $J_{\rm sym}$ on the properties of NSs and HSs,
we show in Fig.~\ref{fig:lambda}
the critical density $\rho_h$ for hyperon appearance in HS matter and the dimensionless tidal deformability $\Lambda_{1.4}$ for NSs and HSs with mass of $1.4M_{\odot}$ by varying individually $K_{\rm sym}$ and $J_{\rm sym}$ in the HSL35 interaction.
It is seen from Fig.~\ref{fig:lambda}~(a)
that a larger value of $K_{\rm sym}$ or $J_{\rm sym}$ leads systematically to a smaller hyperon emergence density, as we have observed and discussed above on the squared sound speed shown in the lowest row of Fig.~\ref{fig:constraints}.
Furthermore,
as depicted in Fig.~\ref{fig:lambda}~(b), it is evident that the $\Lambda_{1.4}$ for both HSs and NSs is in nice agreement with the limit extracted from gravitational wave signal GW170817, namely $190^{+390}_{-120}$~\citep{LIGOScientific:2018cki} as reported by LIGO and Virgo Collaboration.
Note that in Fig.~\ref{fig:lambda}~(b), the $\Lambda_{1.4}$ of HSs is exactly the same as that of NSs, due to the smaller central density in $1.4M_\odot$ HSs below which the hyperons have not yet appear in the HS matter.
Our results therefore indicate that the tidal deformability for both HSs and NSs with HSL35 is nicely compatible with the constraint extracted from gravitational
wave signal GW170817.
In Table~\ref{tab2}, we summarize some basic properties of HSs and NSs for the interactions adopted in Fig.~\ref{fig:constraints}.
It should be noted that as a result of the lower central density of $1.4M_{\odot}$ HSs which has not reached the critical density for hyperon appearance, the quantities corresponding to $1.4M_{\odot}$ HSs are the same as those of $1.4M_{\odot}$ NSs.

%-------------------------------------------------------------------------------------------------------------------

\begin{table*}[t!]
\centering
\caption{The critical density ($\rho_{h}$) for hyperon appearance in NS matter, the dimensionless tidal deformability of $1.4M_{\odot}$ HSs and NSs ($\Lambda_{1.4}$), the central densitiy ($\rho_{\mathrm{cen}}^{1.4}$) and radius ($R_{1.4}$) of $1.4M_{\odot}$ HSs, the central density ($\rho_{\mathrm{cen}}^{\mathrm{TOV}}$), radius ($R_{\mathrm{TOV}}$) and mass ($M_{\mathrm{TOV}}$) of HSs and NSs with the maximum mass, the critical density ($\rho^{\rm DU}_{X}$) and corresponding mass ($M^{\rm DU}_{X}$) of HSs and NSs for the direct Urca process $X$ ($n \rightarrow pl \overline{\nu}_{l}$ in HSs and NSs, $\Lambda \rightarrow pl \overline{\nu}_{l}$ and $\Xi^{-} \rightarrow \Lambda l \overline{\nu}_{l}$ in HSs, with $l$ being electrons or muons) for the interactions shown in Fig.~\ref{fig:constraints}. Note that for some parameter combinations, the direct Urca processes concerning hyperons may have ending points, the densities of which are lower than the central densities of HSs with the maximum mass, while some direct Urca processes about hyperons may not happen at all.}

\scriptsize
\begin{tabular}{cc|c|ccc|cccc|c|c}
\hline\hline
\multicolumn{2}{c|}{} & \multicolumn{1}{c|}{$\rm HSL35$}& \multicolumn{7}{c|}{} & $\rm HSL45$ & $\rm SP6L55$\\
\hline
\multicolumn{2}{c|}{$E_{\rm sym}(\rho_0)(\rm MeV$)} & \multicolumn{1}{c|}{32} & \multicolumn{7}{c|}{$32$} & $34$ & $33$\\
\hline
\multicolumn{2}{c|}{$L(\rm MeV$)} & \multicolumn{1}{c|}{35} & \multicolumn{7}{c|}{$35$} & $45$ & $55$\\
\hline
\multicolumn{2}{c|}{$K_{\rm sym}(\rm MeV$)} & \multicolumn{1}{c|}{$-300$} & \multicolumn{3}{c|}{$-300$} & \multicolumn{4}{c|}{$-210$}& $-315$ & $-100$ \\
\hline
\multicolumn{2}{c|}{$J_{\rm sym}(\rm MeV$)} & $720$ & $760$ & $800$ & $840$ & $720$ & $760$ & $800$ & $840$ & $740$ & $900$\\
\hline
\multicolumn{2}{c|}{$E_{\rm sym}(2\rho_0)(\rm MeV$)} & $31.59$ & $31.88$ & $32.16$ & $32.44$ & $35.66$ & $35.94$ & $36.23$ & $36.51$ & $36.64$ & $50.63$\\
\hline
\multicolumn{2}{c|}{$\rho_{h}(\rho_{0})$} & $3.96$ & $3.88$ & $3.78$ & $3.72$ & $3.92$ & $3.80$ & $3.72$ & $3.64$ & $3.66$ & $3.04$\\
\hline
\multicolumn{2}{c|}{$\Lambda_{1.4}$} & $219.1$ & $233.2$ & $246.5$ & $259.1$ & $242.0$ & $259.5$ & $275.3$ & $289.8$ & $261.1$ & $469.3$\\
\hline
\multicolumn{2}{c|}{$\rho_{\rm cen}^{1.4}(\rho_{0})$} & $3.38$ & $3.29$ & $3.21$ & $3.15$ & $3.43$ & $3.31$ & $3.22$ & $3.14$ & $3.17$ & $2.66$\\
\hline
\multicolumn{2}{c|}{$R_{1.4}(\rm km)$} & $11.38$ & $11.50$ & $11.61$ & $11.71$ & $11.43$ & $11.55$ & $11.65$ & $11.73$ & $12.01$ & $12.79$\\
\hline
\multirow{9}{*}{HS}
& \multicolumn{1}{|c|}{$\rho_{\rm cen}^{\rm TOV}(\rho_{0})$} & $6.28$ & $6.16$ & $6.03$ & $5.94$ & $6.84$ & $6.66$ & $6.47$ & $6.31$ & $6.09$ & $6.06$\\
& \multicolumn{1}{|c|}{$R_{\rm TOV}(\rm km)$} & $11.06$ & $11.17$ & $11.28$ & $11.38$ & $10.66$ & $10.82$ & $10.97$ & $11.10$ & $11.42$ & $11.62$\\
& \multicolumn{1}{|c|}{$M_{\rm TOV}(M_{\odot})$} & $2.03$ & $2.03$ & $2.03$ & $2.03$ & $1.96$ & $1.96$ & $1.96$ & $1.97$ & $2.00$ & $1.90$\\
\cline{2-12}
& \multicolumn{1}{|c|}{$\rho^{\rm DU}_{n \rightarrow pl \overline{\nu}_{l}}(\rho_{0})$} & $4.66$ & $4.46$ & $4.28$ & $4.14$ & $5.10$ & $4.70$ & $4.40$ & $4.16$ & $4.10$ & $2.58$\\
& \multicolumn{1}{|c|}{$M^{\rm DU}_{n \rightarrow pl \overline{\nu}_{l}}(M_{\odot})$} & $1.99$ & $1.98$ & $1.97$ & $1.96$ & $1.90$ & $1.88$ & $1.87$ & $1.86$ & $1.91$ & $1.34$\\
& \multicolumn{1}{|c|}{$\rho^{\rm DU}_{\Lambda \rightarrow pl \overline{\nu}_{l}}(\rho_{0})$} & $4.52$ & $4.30$ & $4.16$ & $4.02$ & --- & $(4.46,5.84)$ & $(4.18,6.22)$ & $4.00$ & $4.04$ & $3.24$\\
& \multicolumn{1}{|c|}{$M^{\rm DU}_{\Lambda \rightarrow pl \overline{\nu}_{l}}(M_{\odot})$} & $1.98$ & $1.97$ & $1.96$ & $1.95$ & --- & $1.86$ & $1.84$ & $1.84$ & $1.91$ & $1.70$\\
& \multicolumn{1}{|c|}{$\rho^{\rm DU}_{\Xi^{-} \rightarrow \Lambda l \overline{\nu}_{l}}(\rho_{0})$} & $4.34$ & $4.16$ & $4.04$ & $3.94$ & $(4.56,5.64)$ & $(4.20,6.32)$ & $4.02$ & $3.88$ & $3.90$ & $3.22$\\
& \multicolumn{1}{|c|}{$M^{\rm DU}_{\Xi^{-} \rightarrow \Lambda l \overline{\nu}_{l}}(M_{\odot})$} & $1.96$ & $1.95$ & $1.95$ & $1.94$ & $1.85$ & $1.82$ & $1.82$ & $1.81$ & $1.88$ & $1.70$\\
\hline
\multirow{5}{*}{NS}
& \multicolumn{1}{|c|}{$\rho_{\rm cen}^{\rm TOV}(\rho_{0})$} & $6.38$ & $6.31$ & $6.22$ & $6.16$ & $6.88$ & $6.69$ & $6.56$ & $6.47$ & $6.22$ & $6.28$\\
& \multicolumn{1}{|c|}{$R_{\rm TOV}(\rm km)$} & $10.68$ & $10.77$ & $10.87$ & $10.95$ & $10.26$ & $10.43$ & $10.56$ & $10.66$ & $10.98$ & $11.06$\\
& \multicolumn{1}{|c|}{$M_{\rm TOV}(M_{\odot})$} & $2.30$ & $2.31$ & $2.30$ & $2.30$ & $2.23$ & $2.24$ & $2.25$ & $2.25$ & $2.29$ & $2.21$\\
\cline{2-12}
& \multicolumn{1}{|c|}{$\rho^{\rm DU}_{n \rightarrow pl \overline{\nu}_{l}}(\rho_{0})$} & $4.74$ & $4.54$ & $4.36$ & $4.20$ & $5.32$ & $4.84$ & $4.50$ & $4.24$ & $4.14$ & $2.58$\\
& \multicolumn{1}{|c|}{$M^{\rm DU}_{n \rightarrow pl \overline{\nu}_{l}}(M_{\odot})$} & $2.19$ & $2.17$ & $2.15$ & $2.12$ & $2.15$ & $2.10$ & $2.06$ & $2.02$ & $2.07$ & $1.34$\\
\hline
\hline
\end{tabular}
\label{tab2}
\end{table*}
%-------------------------------------------------------------------------------------------------------------------

\begin{figure}[htp!]
\centering
\includegraphics[width=0.95\columnwidth]{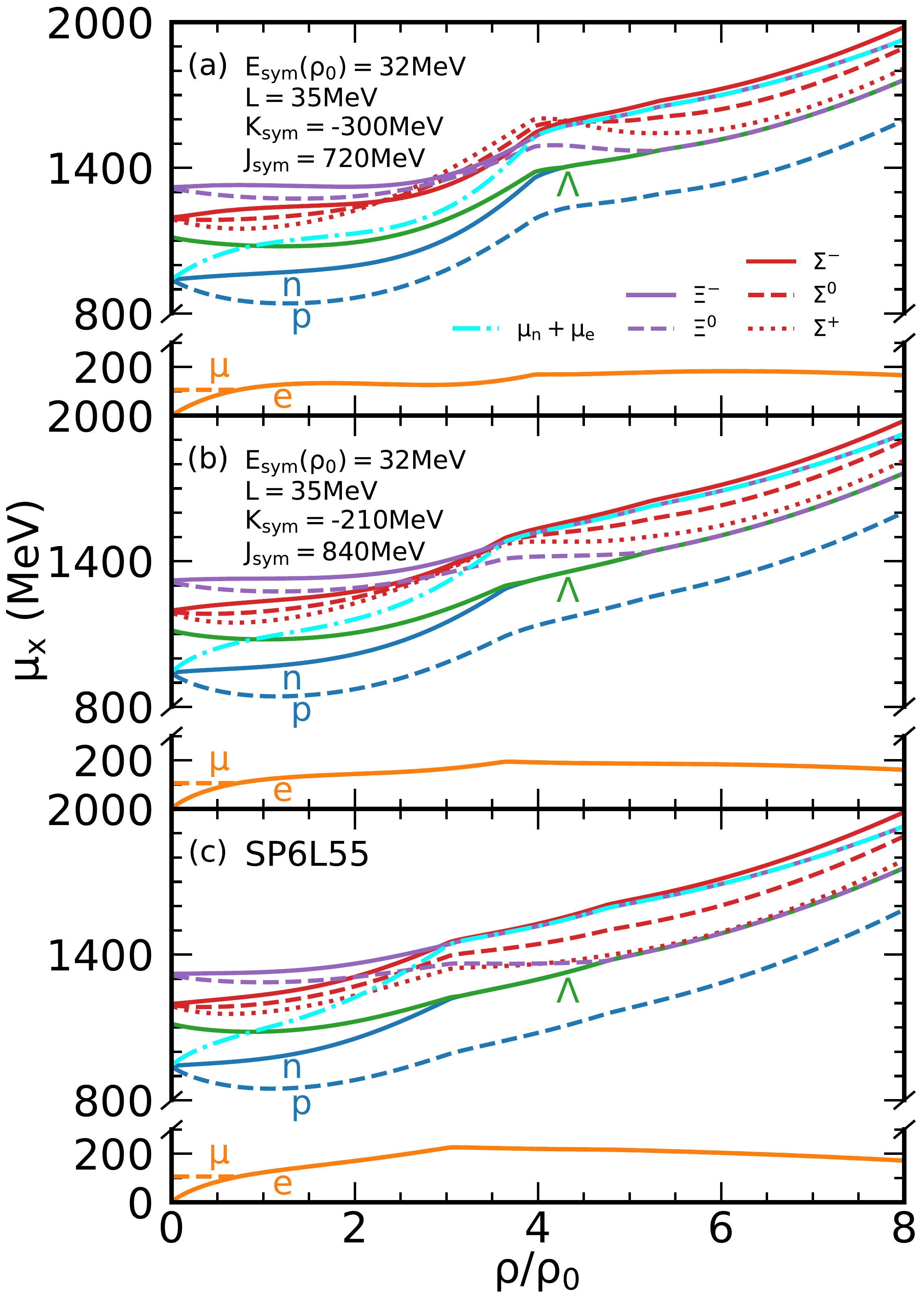}
\caption{Particle chemical potentials in HS matter as a function of baryon density for the HSL35 interaction (a),  the HSL35 but with $K_{\rm sym} = -210$~MeV and $J_{\rm sym} = 840$~MeV (b), and the SP6L55 interaction (c). The chemical potential sum $\mu_n + \mu_e$ for neutrons
and electrons is also included.}
\label{fig:ChemPot}
\end{figure}

It is instructive to have a further comprehensive understanding of the influence of $K_{\rm sym}$ and $J_{\rm sym}$ on properties of NSs and HSs.
From Fig.~\ref{fig:constraints}~(c) and (d), it is evident that a smaller value of $K_{\rm sym}$ combined with a smaller value of $J_{\rm sym}$ leads to a decrease of $E_{\rm sym}(\rho)$ with density around $2-3\rho_0$.
Consequently, the chemical potentials of neutrons in nuclear matter at these densities are reduced accordingly, making it hard for neutrons decaying into hyperons and thus pushing the hyperon appearance density to a higher value.
On the other hand, beyond a density higher than about $4\rho_0$, there is an enhancement in $E_{\rm sym}(\rho)$ which increases the neutron chemical potential and favors the appearance of hyperons.
In other words, having a soft density behavior for $E_{\rm sym}(\rho)$ at densities between $2-3\rho_0$ but stiff above $4\rho_0$ can delay the critical emergence density for hyperons.
This can be seen more clearly from Fig.~\ref{fig:ChemPot} which shows the density dependence of particle chemical potentials in HS matter by varying individually $K_{\rm sym}$ and $J_{\rm sym}$ in the HSL35 interaction. For comparison, the corresponding results with SP6L55 are also included in Fig.~\ref{fig:ChemPot}.
In particular,
we note that the neutron chemical potential at $\rho = 2(3)\rho_0$ is $997(1107)$~MeV for $K_{\rm sym} = -300$~MeV and $J_{\rm sym} = 720$~MeV (a), $1017(1149)$~MeV for $K_{\rm sym} = -210$~MeV and $J_{\rm sym} = 840$~MeV (b), while it becomes $1054(1211)$~MeV for SP6L55 which comparatively has a much stiffer symmetry energy at intermediate density.
Although both parameters $K_{\rm sym}$ and $J_{\rm sym}$ affect the critical emergence density of hyperons, their respective effects on the mass-radius relation and $M_{\rm TOV}$ are different due to their distinct impacts on the EOS for HS matter as reflected by the variations in the squared sound speed shown in the lowest row of Fig.~\ref{fig:constraints}.
Therefore, the combined influence of $K_{\rm sym}$ and $J_{\rm sym}$ on the critical density for hyperon appearance and overall density behavior of HS matter EOS collectively shapes the properties of HSs.

\begin{figure}[htp!]
\centering
\includegraphics[width=0.98\columnwidth]{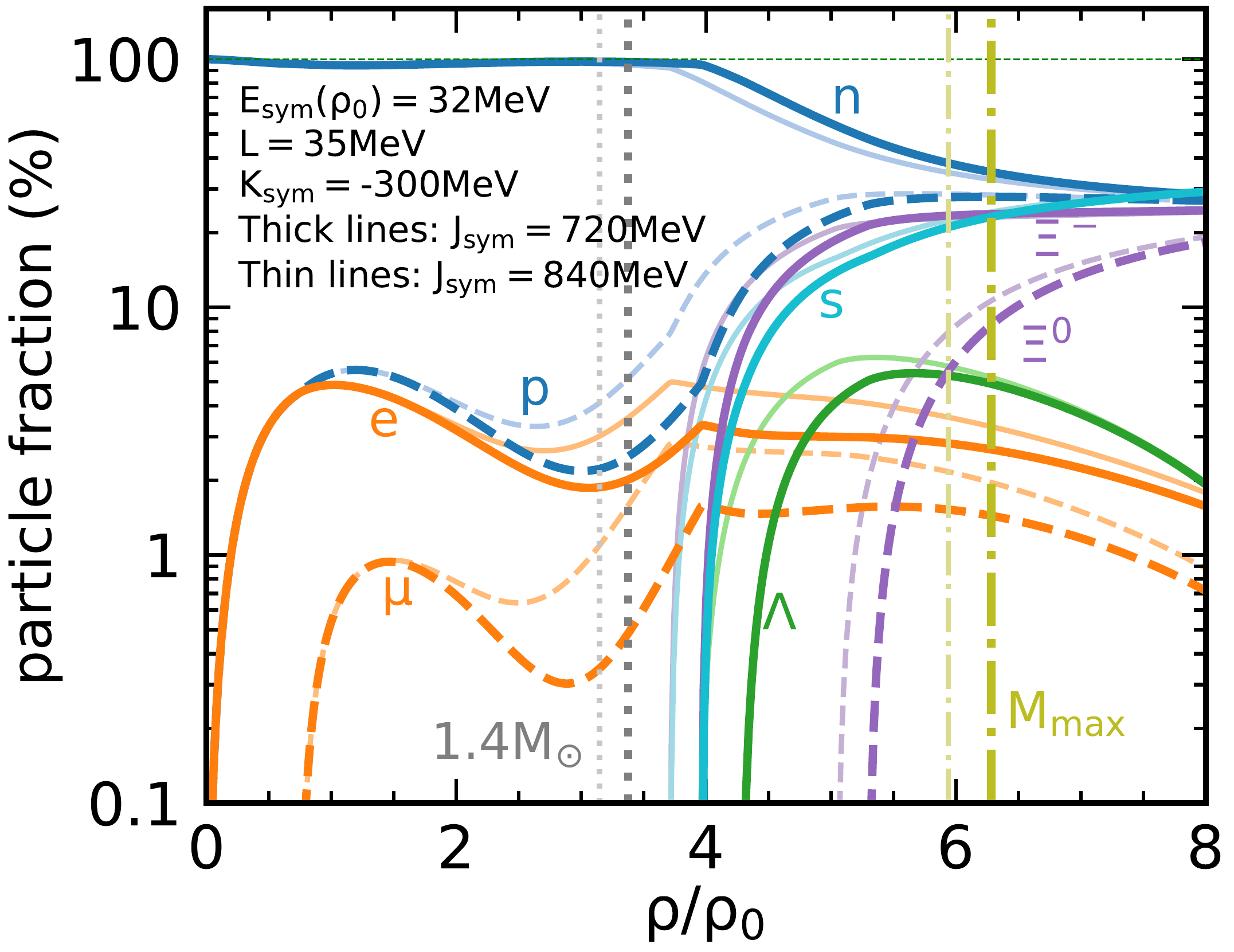}
\caption{Particle fractions in HS matter as a function of baryon density for the HSL35 interaction but with $J_{\rm sym}=720$~MeV (thick lines) and $840$~MeV (thin lines).
Additionally, the central density for HSs with mass of $1.4 M_{\odot}$ and maximum mass is indicated by the vertical lines, and the strangeness fraction denoted by ``s" is also included.}
\label{fig:ratio}
\end{figure}

Finally, we show in Fig.~\ref{fig:ratio} the particle fraction and the strangeness fraction $\frac{\rho_s}{3\rho_B}$ in HS matter, by varying $J_{\rm sym}$ from $720$~MeV to $840$~MeV in HSL35. Here, $\rho_s$ represents the total number density of constituent strange quarks within hadrons.
Fig.~\ref{fig:ratio}
provides a clear depiction of the impact of $J_{\rm sym}$ on the critical emergence density of hyperons, i.e., a larger $J_{\rm sym}$ leads to a smaller critical emergence density of hyperons. In particular, $\Xi^-$ appear first, then followed by $\Lambda $ and last $\Xi^0$ as the density increases.
The earlier appearance of $\Xi^-$ is mainly because of the larger value of chemical potential sum $\mu_n+\mu_e$ for neutrons and electrons (see Fig.~\ref{fig:ChemPot}), which triggers the process $n+e^{-} \rightarrow \Xi^- + \nu_e$.
In addition,
it is seen from Fig.~\ref{fig:ratio} that $\Sigma$ hyperons do not appear at densities below $8\rho_0$ in the parameter space considered here, mainly due to their repulsive potentials which lead to large chemical potentials for $\Sigma$ hyperons as shown in Fig.~\ref{fig:ChemPot}.
Furthermore, one can see clearly that the strangeness is not present for NSs with mass smaller than $1.4 M_{\odot}$ and the strangeness fraction can reach to about $25\%$ in the center of HSs with maximum mass configuration.
Notably, it can be observed that the densities corresponding to the peaks (including the minor peaks) in $C^2_s$ for HS matter (as shown in the lowest row of Fig.~\ref{fig:constraints}) exactly coincide with the emergence densities of various types of hyperons.

%-------------------------------------------------------------------------------------------------------------------
%-------------------------------------------------------------------------------------------------------------------

\subsection{Implications of a soft symmetry energy at intermediate densities}
It is interesting to see that the HSL35 predicts a symmetry energy with a negative slope between $\rho_0$ and $3\rho_0$ followed by a steep increase. We note this feature is in nice agreement with the recent findings in the relativistic mean-field (RMF) model with isovector-scalar $\delta$ meson and its coupling to isoscalar-scalar $\sigma$ meson, which indicate that the $\delta$-$\sigma$ mixing can soften the symmetry energy $E_{\rm sym}(\rho)$ at intermediate densities while stiffen the $E_{\rm sym}(\rho)$ at high densities~\citep{Zabari:2018tjk,Zabari:2019clk,Kubis:2020ysv,Miyatsu:2022wuy,Li:2022okx,Miyatsu:2023lki}. This interesting feature for the symmetry energy with the $\delta$-$\sigma$ mixing is important to remove the tension between PREX-II and GW170817 observed in the conventional RMF model~\citep{Li:2022okx}. In addition, the negative slope at suprasaturation densities is actually also consistent with the constraints~\citep{Xiao:2008vm} from analyzing charged pion ratio in heavy-ion collisions at FOPI~\citep{FOPI:2006ifg} although somewhat the model dependence still exists, see, e.g., Refs.~\citep{Feng:2009am,Xie:2013np,Xu:2013aza,Hong:2013yva,Song:2015hua,Zhang:2017mps,Zhang:2018ool}.

The symmetry energy with a negative slope between $\rho_0$ and $3\rho_0$ followed by a steep increase will lead to special density dependence of proton fraction and thus influence the occurrence of the direct Urca~(DU) process, which can enhance the emission of neutrinos and make it a more important
process in the cooling of a NS~\citep{Lattimer:1991ib}.
In Table~\ref{tab2},
we show the critical density and the corresponding mass of HSs and NSs for the DU processes $n \rightarrow p+l+\overline{\nu}_{l}$ and $p+l \rightarrow n+\nu_{l}$ in HSs and NSs, $\Lambda \rightarrow p+l+\overline{\nu}_{l}$, $p+l \rightarrow \Lambda+\nu_{l}$, $\Xi^{-} \rightarrow \Lambda+l+\overline{\nu}_{l}$ and $\Lambda+l \rightarrow \Xi^{-}+\nu_{l}$ in HSs with leptons $l$ being electrons or muons.
It is interesting to see that for certain parameter combinations, e.g., $E_{\rm sym}(\rho_0) = 32$ MeV, $L = 35$ MeV, $K_{\rm sym} = -210$~MeV and $J_{\rm sym} = 720, 760, 800$~MeV, the DU processes concerning hyperons may end at densities lower than the central densities of HSs with maximum mass configuration, while some DU processes about hyperons, e.g., $\Lambda \rightarrow p+l+\overline{\nu}_{l}$ and $p+l \rightarrow \Lambda+\nu_{l}$ for $E_{\rm sym}(\rho_0) = 32$ MeV, $L = 35$ MeV, $K_{\rm sym} = -210$~MeV and $J_{\rm sym} = 720$~MeV, may not happen at all.
Indeed, the soft symmetry energy at intermediate densities (e.g., HSL35 and HSL45) leads to higher critical density and thus larger critical compact star mass for the DU process, i.e., $1.96M_{\odot}$ for HSs and $2.19M_{\odot}$ for NSs with HSL35 while $1.88M_{\odot}$ for HSs and $2.07M_{\odot}$ for NSs with HSL45,
compared to the case with SP6L55 which predicts a critical mass of $1.34M_{\odot}$ for the DU process for both NSs and HSs.
These results imply that the DU processes are relatively hard to occur in NSs and HSs for the soft symmetry energy at intermediate densities (e.g., HSL35 and HSL45), and the larger critical mass for the DU process with HSL35 seems to support the finding that the DU process shall not occur in typical
NSs with masses in the range of $1.0-1.5M_{\odot}$ obtained from population syntheses scenarios~\citep{Blaschke:2004vq,Popov:2004ey,Klahn:2006ir}.
Nevertheless, it should be mentioned that the cooling of NSs and HSs can
be significantly influenced by nucleon pairing effects which are still largely uncertain~\citep{Page:2009fu,Potekhin:2019eya}.

%----------------------------------------------------------------------------------------------------
\section{Conclusion and outlook}
\label{Sec:Summary}
We have extended the nuclear N3LO Skyrme pseudopotential to include the interactions of octet baryons
by assuming that the hyperon-nucleon and hyperon-hyperon interactions have
similar density, momentum, and isospin dependence as for the nucleon-nucleon interaction. The parameters
in these interactions are determined from either experimental information if any or microscopic calculations.
By considering
the constraints from microscopic calculations, terrestrial experiments and astrophysical observations,
we have demonstrated that an appropriate high density behavior of the symmetry energy, namely, soft around $2-3\rho_0$ but stiff above about $4\rho_0$, offers a potential solution to the hyperon puzzle.
The softening of the symmetry energy around $2-3\rho_0$ can effectively enhance the critical density for hyperon appearance in hyperon stars and thus mitigate the hyperon influence on properties of hyperon stars.
In addition,
the squared sound speed of the hyperon star matter exhibits a strong peak structure at
density of $3-4\rho_0$ when the hyperons start to appear, providing a natural explanation on the similar phenomenon observed in the model-independent analysis on the multimessenger data.

In particular, our results indicate that the extended N3LO Skyrme pseudopotential HSL35 with a symmetry energy
given by $E_{\rm sym}(\rho_0) = 32$ MeV, slope parameter $L = 35$ MeV, curvature parameter $K_{\rm sym} = -300$ MeV and skewness parameter $J_{\rm sym} = 720$ MeV, which is soft around $2-3\rho_0$ but stiff above about $4\rho_0$, is able to predict a maximum mass of $M_{\rm TOV} \gtrsim 2M_\odot$ for static hyperon stars. At the same time, this extended N3LO Skyrme pseudopotential can be compatible with various constraints from theory, experiments and observations, including microscopic calculations on pure neutron matter, flow data in heavy-ion collisions, the tidal deformability $\Lambda_{1.4}$ from gravitational wave event GW170817, the observed mass-radius relation for compact stars from NICER as well as the unusually low mass and small radius in the central compact object of HESS J1731-347.

In future, it will be interesting to incorporate the extended N3LO Skyrme pseudopotential for the octet baryons in nuclear structure calculations and the transport model simulations for heavy-ion collisions. With the ongoing accumulation of more and more high quality data of hypernuclei as well as hyperon/hypernuclei production in heavy-ion collisions in terrestrial laboratories, more stringent constraints shall be obtained on the effective hyperon-nucleon and hyperon-hyperon interactions. These investigations will be useful to further address more precisely the hyperon puzzle as well as the hyperon roles on the dynamics of supernova explosions and binary neutron star mergers, and eventually to shed light on the QCD phase diagram of baryon-rich matter.

%----------------------------------------------------------------------------------------------------
\section*{Acknowledgements}
This work was supported in part by the National SKA Program of China (Grant No. 2020SKA0120300), the National Natural Science Foundation of China under Grant No 12235010, and the Science and Technology Commission of Shanghai Municipality (Grant No. 23JC1402700).


\bibliography{RefHypStarEsym}{}
\bibliographystyle{aasjournal}



\end{document}